\documentstyle[amsfonts,thmsc,11pt,sw20lart]{article}

\def\ss{\relax{\supset\kern -13pt +}}
 \makeatletter
 \@addtoreset{equation}{section}
 \makeatother
\input{tcilatex}
\begin{document}

\title{Umbral Calculus, Difference Equations and the Discrete Schr\"{o}dinger
Equation}
\author{Decio Levi \\
Dipartimento di Fisica, Universit\`{a} di Roma Tre and INFN, \\
Sezione di Roma Tre, via della Vasca Navale 84, Roma, Italy \\
e-mail address: levi@fis.uniroma3.it \and Piergiulio Tempesta \\
Centre de recherches math\'{e}matiques, Universit\'{e} de Montr\'{e}al, C.P.
6128, \\
succ. Centre-ville, H3C 3J7, Montr\'{e}al (Qu\'{e}bec), Canada.\\
e-mail address: tempesta@crm.umontreal.ca \and Pavel Winternitz \\
Centre de recherches math\'{e}matiques and D\'{e}partement de
math\'{e}matiques \\
et de statistique, Universit\'{e} de Montr\'{e}al, C.P. 6128, \\
succ. Centre-ville, H3C 3J7, Montr\'{e}al (Qu\'{e}bec), Canada\\
e-mail address: wintern@crm.umontreal.ca}
\maketitle

\begin{abstract}
We discuss umbral calculus as a method of systematically discretizing linear
differential equations while preserving their point symmetries as well as
generalized symmetries. The method is then applied to the Schr\"{o}dinger
equation in order to obtain a realization of nonrelativistic quantum
mechanics in discrete space--time. In this approach a quantum system on a
lattice has a symmetry algebra isomorphic to that of the continuous case.
Moreover, systems that are integrable, superintegrable or exactly solvable
preserve these properties in the discrete case.
\end{abstract}

\renewcommand{\theequation}{\thesection.\arabic{equation}} 


PACS num. 02.20.--a, 02.30.Ik, 02.30.Ks, 03.

\section{Introduction}

A sizable literature exists on discrete quantum mechanics, that is on
quantum mechanics in discrete space--time. We refer to a recent review for
motivation and for an extensive list of references \cite{Gibbs}. There are
many reasons for considering quantum systems in discrete space--time. One is
that physical space--time may indeed be discrete, involving an elementary
length and some minimal time interval. Then continuous theories would only
be approximations to the real world. Another reason is the usual one: on a
lattice one can avoid some of the divergence problems occuring in quantum
field theories. On the other hand, some properties of quantum systems are
lost in any discretization. The aim of this article is to discuss a
discretization of space--time in which the Schr\"{o}dinger equation is
replaced by a difference equation. This is done in such a manner that many
of the essential properties of the continuous system are preserved. In
particular, we preserve the group theoretical and integrability properties
of the Schr\"{o}dinger equation. This is true for the time--dependent, as
well as the stationary equation. The free equations, as well as those with
potentials, after discretization have symmetry groups, isomorphic to those
of the continuous case. Lie point symmetries, after discretization, may
however act at several points of the lattice.

Another property that we wish to preserve is that of integrability, and also
superintegrability. By integrability, for an $n$ dimensional quantum system,
we mean the existence of $n$ well defined algebraically independent
Hermitian operators $\left\{ X_{1},...,X_{n}\right\} $ (including the
Hamiltonian $H$) commuting pairwise. Superintegrability means that there
exist further independent operators, $\left\{ Y_{1},...,Y_{k}\right\} ,$ $%
1\leq k\leq n-1,\,$commuting with the Hamiltonian $H$, but not necessarily
with the other operators $X_{i},$ nor $Y_{i}$ \cite{FMUSW}--\cite{KKMP}.

Finally, we wish to preserve exact solvability in the discretization, i.e.
the fact that for certain systems (like the harmonic oscillator, or hydrogen
atom) it is possible to calculate all energy levels algebraically.

A mathematical tool that we shall use for the study of symmetries and exact
solutions of linear equations is the so--called ''umbral calculus''. This
calculus, which originated in 19th century with the work of Sylvester,
Cayley and others, was used for a long time as a useful tool to derive
combinatorial identities (see, for instance, \cite{Riord}). Nevertheless, it
was only with Rota et al. \cite{RKO}--\cite{Roman} that this calculus was
put on an axiomatic basis using the language of linear algebra of operators.
In Ref. \cite{BL}, the interested reader can find an up to date survey
concerning the origins of umbral calculus and its many applications in
several branches of mathematics, like combinatorics, functional analysis,
algebraic topology, theory of special functions and orthogonal polynomials,
etc.

Umbral calculus has recently been used explicitly \cite{DMS}, or implicitly 
\cite{GS}--\cite{CT}, to provide discrete representations of canonical
commutation relations, specially in the context of exactly solvable and
quasi exactly solvable quantum systems \cite{T2}--\cite{T3}. Linear
differential equations have been discretized in a symmetry preserving manner
using commuting difference operators \cite{FNNV}--\cite{FV}. An alternative
approach \cite{LVW} to symmetries of linear difference equations makes use
of a discretized version of the prolongation theory of evolutionary vector
fields. Finally, umbral calculus was used in an implicit manner, to obtain
several different symmetry preserving discretizations of the linear heat
equation \cite{LND}.

Symmetries of difference equations, mainly nonlinear ones, have recently
received a lot of attention (see e.g. \cite{FNNV}--\cite{DKW} and references
therein). What has emerged for purely difference equations is that in order
to capture the essential features and usefulness of symmetries of
differential equations it is necessary to make serious adjustments. Either
one must go beyond point transformations to generalized ones \cite{RLW}--%
\cite{RLRW}, or one must use symmetry adapted and transforming lattices (as
proposed initially by Dorodnitsyn) \cite{LTW1}--\cite{DKW}.

In this paper we follow the first approach. We consider a fixed lattice and
use umbral calculus to obtain symmetries acting simultaneously on more than
one point of the lattice. We apply this approach to quantum mechanics.

\section{\protect\smallskip Umbral Calculus}

To make this article self--contained let us sum up in Section 2.1 some known
definitions and also some results proven as theorems by Rota \cite{Rota},
Roman \cite{Roman} and Dimakis et al. \cite{DMS}.

We shall actually need umbral calculus on spaces of many variables in order
to study multidimensional difference equations. However, for simplicity of
exposition and notation we shall in this section restrict to the case of one
variable $x$.

\subsection{General Theory.}

Let ${\frak {F}}$ be the algebra of formal power series in a variable $x$,
and ${\cal P}$ the algebra of polynomials in the same variable. The algebras
will be considered over a field ${\Bbb {F}}$\thinspace of characteristic
zero. This field in the subsequent considerations will be identified with $%
{\Bbb {R} }$ or ${\Bbb {C}}$. An element of ${\frak {F}}$ is of the form

\begin{equation}
\sum_{k}a_{k}x^{k}\equiv f\left( x\right) .  \label{2.1}
\end{equation}
The operations defined in ${\frak {F}}$ are the addition of series 
\begin{equation}
\sum_{k=0}^{\infty }a_{k}x^{k}+\sum_{k=0}^{\infty
}b_{k}x^{k}=\sum_{k=0}^{\infty }\left( a_{k}+b_{k}\right) x^{k},  \label{2.2}
\end{equation}
and the multiplication 
\begin{equation}
\left( \sum_{k=0}^{\infty }a_{k}x^{k}\right) \left( \sum_{l=0}^{\infty
}b_{l}x^{l}\right) =\sum_{m=0}^{\infty }\left(
\sum_{j=0}^{m}a_{j}b_{m-j}\right) x^{m}.  \label{2.3}
\end{equation}
The algebra ${\frak {F}}$ \thinspace is also called the {\it umbral algebra} 
\cite{Roman}.

A polynomial sequence $p_{n}\left( x\right) \in {\cal P}$ is a sequence
whose $n$--th element is a polynomial of degree $n$. We will denote by $%
{\cal L}$ the algebra of linear operators acting on ${\frak {F}}$ or ${\cal P%
}. $

\begin{definition}
A shift operator $T$ $\in {\cal L}$ $\,$is a linear operator such that 
\begin{equation}
T\,p\left( x\right) =p\left( x+\sigma \right) ,  \label{2.4}
\end{equation}
where $p(x)$ is a polynomial and $\sigma \in {\Bbb {F}}$.
\end{definition}

\begin{definition}
An operator $F\in {\cal L}$ is said to be shift--invariant if it commutes
with all shift operators (i.e. with $T$ for all values of $\sigma $).
\end{definition}

\begin{definition}
An operator $U$ is said to be a {\it delta operator} if it is
shift--invariant and 
\begin{equation}
U\,x=c\neq 0,  \label{2.5}
\end{equation}
where $c\in {\Bbb {F}}$.
\end{definition}

Using Definition 2.1 and the linearity of shift--invariant operators one can
prove the following result. If $U$ is a delta operator, for every $c\in 
{\Bbb {F}}$ we have 
\begin{equation}
U\,c=0.  \label{2.6}
\end{equation}

\begin{definition}
A polynomial sequence $p_{n}\left( x\right) ,$ $n=0,1,2,...,$ is called a
sequence of basic polynomials for the delta operator $U$ if 
\begin{equation}
p_{0}\left( x\right) =1,\quad p_{n}\left( 0\right) =0\quad \forall n>0,
\label{2.7}
\end{equation}
and 
\begin{equation}
U\,\,p_{n}\left( x\right) =n\,p_{n-1}\left( x\right) .  \label{2.8}
\end{equation}
\end{definition}

It is easy to show that every delta operator has a unique sequence of basic
polynomials. We will denote by ${\frak {I}}$ the one--to--one correspondence
between basic sequences and delta operators.

Let ${\cal A}$ be the algebra of shift--invariant operators, endowed with
the usual operations of sum of two operators, product of a scalar with an
operator, and product of two operators. We introduce a multiplication
operation $*:$ ${\cal A\times L}\rightarrow {\cal L}$, defined by 
\begin{equation}
F*O=\left[ F,O\right] =FO-OF,  \label{2.9}
\end{equation}
where $F$ is a shift--invariant operator and $O\in {\cal L}$. In particular,
if $x\,\,$denotes the multiplication operator $x:p\left( x\right)
\rightarrow x\,p\left( x\right) ,$ then $F*x$ corresponds to what in the
umbral literature is known as the Pincherle derivative of $F$. In this case
we will write 
\begin{equation}
F^{\prime }=F*x=\left[ F,x\right] .  \label{2.11}
\end{equation}
Using the $*$ multiplication the Leibnitz rule becomes 
\begin{equation}
F*\left( fg\right) =\left( F*f\right) g+f\left( F*g\right)  \label{2.12}
\end{equation}
and the Jacobi identity is expressed by 
\begin{equation}
F*G*H+G*H*F+H*F*G=0.  \label{2.13}
\end{equation}
Let us now consider a pair of shift--invariant operators: a {\it delta}
operator $U\in {\cal L}$ and its conjugate operator $\beta \in {\cal L}$,
defined in such a way that the {\it Heisenberg--Weyl algebra} is satisfied: 
\begin{equation}
\left[ U,\text{ }x\,\beta \right] =1.  \label{2.14}
\end{equation}
It is possible to prove that, if $U$ is a delta operator, then the inverse
of $U^{\prime }\,$exists (\cite{Rota}, p. 18). Therefore, the operator $%
\beta \,$is determined by the relation: 
\begin{equation}
\beta =\left( U^{\prime }\right) ^{-1}.  \label{2.15}
\end{equation}
To prove this it suffices to notice that 
\[
1=\left[ U,\text{ }x\,\beta \right] =\left[ U,x\right] \,\beta =U^{\prime
}\,\beta , 
\]
where the property $\left[ U,\beta \right] =0$ has been exploited. Eq.(\ref
{2.15}) follows.

Let us present some specific examples of realizations of the conjugate
operators $U$ and $\beta $ in terms of derivatives and shifts, respectively.

{\bf Example 2.1}. The continuous case. We have: 
\begin{equation}
U=\partial _{x},\quad \beta =1.  \label{2.15b}
\end{equation}

{\bf Example 2.2}. The discrete case. The variable $x\,$ is defined over an
equally spaced lattice, with spacing $\sigma $. Two of the most common
choices for the discrete derivative are:

a) The right discrete derivative. 
\begin{equation}
U=\Delta ^{+}=\frac{T-1}{\sigma },\quad \beta =T^{-1}.  \label{2.16}
\end{equation}

b) The left discrete derivative. 
\begin{equation}
U=\Delta ^{-}=\frac{1-T^{-1}}{\sigma },\quad \beta =T.  \label{2.17}
\end{equation}
Other cases will be considered below.

Using the $*$ multiplication of eq. (\ref{2.9}) it is easy to construct the
basic sequence for the operator $U$. Let us introduce the polynomial
sequence of operators

\begin{equation}
P_{n}\,=\left( x\beta \right) ^{n},\qquad n\in {\Bbb {N}.}  \label{2.21}
\end{equation}
The delta operator $U$ satisfies the relation

\begin{equation}
\left[ U,\,\left( x\beta \right) ^{n}\right] =n\left( x\beta \right)
^{n-1},\qquad n\in {\Bbb {N}.}  \label{2.22}
\end{equation}
This is an immediate consequence of the definition (\ref{2.14}) and of the
Leibnitz rule (\ref{2.12}). A proof is obtained by induction. From (\ref
{2.22}) we immediately obtain: 
\begin{equation}
U*P_{n}=nP_{n-1},\qquad n\in {\Bbb {N}}  \label{2.23}
\end{equation}
This shows that $\left\{ \left( x\beta \right) ^{n}\right\} _{n\in {\Bbb {N}}%
}$ is the basic sequence for the operator $U$, under the $*$ multiplication.

\begin{definition}
An umbral correspondence is a map ${\cal R}:$ ${\cal L}\rightarrow {\cal L}$
defined by 
\begin{equation}
\left( x\beta _{1}\right) ^{n}\stackrel{\cal R}{\rightarrow \,}\left( x\beta
_{2}\right) ^{n},  \label{2.24}
\end{equation}
where $P_{n}^{1}=\left\{ \left( x\beta _{1}\right) ^{n}\right\} $ and $%
P_{n}^{2}=\left\{ \left( x\beta _{2}\right) ^{n}\right\} \,$are basic
sequences of operators for two delta operators $U_{1}$ and $U_{2}$
respectively.
\end{definition}

The umbral correspondence (\ref{2.24}) naturally induces a correspondence
between the two operators $U_{1}$ and $U_{2}$, according to the following
scheme:
\[
\left( x\beta _{1}\right) ^{n}\stackrel{\cal R}{\leftrightarrow \,}\left(
x\beta _{2}\right) ^{n}
\]
\begin{equation}
{\frak I}\updownarrow \quad \quad \quad \quad {\frak I}\updownarrow 
\label{2.25}
\end{equation}
\[
U_{1}\quad \stackrel{\cal R}{\leftrightarrow \,}\quad U_{2}.
\]
We shall also denote the induced correspondence between delta operators by
the symbol ${\cal R}$.

Systems of equations connected by the umbral map (\ref{2.24}) share many
algebraic properties. A particular case of the umbral correspondence is when 
$U_{1}$ is the standard derivative $\partial _{x}$, and $U_{2}$ is a
discrete derivative $\Delta $. Then according to the scheme (\ref{2.25}) we
have 
\begin{equation}
x^{n}\stackrel{\cal R}{\quad \leftrightarrow \,}\left( x\beta \right) ^{n}
\label{2.26}
\end{equation}
\[
{\frak {I}}\updownarrow {\frak \quad \quad \quad {I}}\updownarrow 
\]
\begin{equation}
\partial _{x}\stackrel{\cal R}{\quad \leftrightarrow \,}\Delta   \label{2.27}
\end{equation}

Let us observe that Definition 2.5 generalizes the notion of an umbral
operator introduced in \cite{Rota}: an umbral operator $R$ $:{\cal %
P\rightarrow P}$ is an operator (in general not necessarily
shift--invariant) which maps some basic sequence of polynomials $p_{n}\left(
x\right) $ into another basic sequence $q_{n}\left( x\right) :$%
\begin{equation}
\,p_{n}\left( x\right) \stackrel{R}{\leftrightarrow }q_{n}\left( x\right) .
\label{2.28}
\end{equation}
Indeed, we observe that, since $\beta $ is a function of shifts and any
constant is invariant under the action of a shift operator, an umbral
operator $R$ is deduced from the action of ${\cal R}$ simply applying the
sequence of operators $\left( x\beta \right) ^{n}$ onto $1$: 
\begin{equation}
\left( x\beta _{1}\right) ^{n}\cdot 1\stackrel{R}{\leftrightarrow }\left(
x\beta _{2}\right) ^{n}\cdot 1.  \label{2.29}
\end{equation}
From (\ref{2.22}) we also get 
\begin{equation}
U_{i}\left( x\beta _{i}\right) ^{n}\cdot 1=n\,\left( x\beta _{i}\right)
^{n-1}\cdot 1,\qquad i=1,2.  \label{2.30}
\end{equation}

An important consequence is that the umbral correspondence (\ref{2.25})\
preserves commutation relations between operators in ${\cal L}$. In
particular, it preserves Lie algebras.

Indeed, let $A_{1}$ be a $m$--dimensional Lie algebra, generated by vector
fields $\left\{ {\bf v}_{i}\right\} _{i=1,...,m}$ of the form 
\begin{equation}
{\bf v}_{i}=\sum_{j}a_{j}\left( x_{1},..,x_{p}\right) \partial _{x_{j}}.
\label{2.31}
\end{equation}
The umbral correspondence (\ref{2.25}) maps $A_{1}$ isomorphically into an
algebra $A_{2}$, generated by the vector fields $\left\{ {\bf v}%
_{i}^{U}\right\} _{i=1,...,m},$ with 
\begin{equation}
{\bf v}_{i}^{U}=\sum_{j}a_{j}\left( x_{1}\beta _{x_{1}},...,x_{p}\beta
_{x_{p}}\right) \Delta _{x_{j}}.  \label{2.32}
\end{equation}
\newline
This follows from the fact that the umbral correspondence (\ref{2.25})
preserves the Heisenberg--Weyl algebra.

\subsection{Umbral Calculus and Linear Difference Operators}

The umbral approach reveals its power in the study of linear difference
operators.

For our purposes, namely the study of difference equations and their
continuous limits, we shall need only two types of delta operators. The
first is simply the derivative $U=\partial _{x}$, with $\beta =1$. The
second is a general difference operator that has $\partial _{x}$ as its
continuous limit. We put 
\begin{equation}
U\equiv \Delta =\frac{1}{\sigma }\sum_{k=l}^{m}a_{k}T_{\sigma }^{k},\qquad
l,m\in {\Bbb {Z},\qquad }l<m  \label{2.35}
\end{equation}
where $a_{k}\,$and $\sigma $ are constants and $T_{\sigma }\equiv T\,$is the
shift operator of eq.(\ref{2.4}). In order for $\Delta $ in (\ref{2.35}) to
be a delta operator, it must satisfy eq.(\ref{2.5}). For any function $%
f\left( x\right) \in {\frak {F}}$ eq. (\ref{2.35}) implies 
\begin{equation}
\Delta f\left( x\right) =\frac{1}{\sigma }\sum_{k=l}^{m}a_{k}T^{k}f\left(
x\right) =\frac{1}{\sigma }\sum_{k=l}^{m}a_{k}f\left( x+k\sigma \right) .
\label{2.36}
\end{equation}
Using a Taylor expansion around $\sigma =0$ we get 
\begin{equation}
\Delta f\left( x\right) =\frac{1}{\sigma }\sum_{q=0}^{\infty }\frac{%
f^{\left( q\right) }\left( x\right) }{q!}\sigma ^{q}\sum_{k=l}^{m}a_{k}k^{q}
\label{2.37}
\end{equation}
Choosing $f\left( x\right) =x\,$we immediately see that eq. (\ref{2.5})
implies 
\begin{equation}
\sum_{k=l}^{m}a_{k}=0,  \label{2.38}
\end{equation}
and $\sum_{k=l}^{m}a_{k}k=c$. We require that in the continuous limit $%
\Delta \,$be the derivative $\partial _{x}$; this implies $c=1,$ i.e. 
\begin{equation}
\sum_{k=l}^{m}a_{k}k=1.  \label{2.38b}
\end{equation}
Eq. (\ref{2.35}) involves $m-l+1\,$constants $a_{k}$, subject to two
conditions (\ref{2.38}) and (\ref{2.38b}). To fix all constants $a_{k}\,$we
must impose $m-l-1\,$further conditions, for instance 
\begin{equation}
\gamma _{q}\equiv \sum_{k=l}^{m}a_{k}k^{q}=0,\qquad q=2,3,...,m-l.
\label{2.39}
\end{equation}
Conditions (\ref{2.38}) and (\ref{2.38b}) are necessary and sufficient for $%
U=\Delta $ to be a delta operator which has the derivative $\partial _{x}$
as its continuous limit.

\begin{definition}
A difference operator of order $p=m-l\,$is a delta operator of the form (\ref
{2.35}) satisfying eqs. (\ref{2.38}) and (\ref{2.38b}).
\end{definition}

\begin{theorem}
If the difference operator $\Delta $ of order ($m-l)\geq 2$ satisfies the
supplementary conditions (\ref{2.39}) it provides an approximation of order $%
\sigma ^{m-l}$ of the derivative $\partial _{x}$.
\end{theorem}

{\bf Proof.} We immediately have from eq. (\ref{2.37}) and eqs. (\ref{2.38}%
), (\ref{2.38b}) and (\ref{2.39}) 
\begin{equation}
\Delta f\stackunder{\sigma \rightarrow 0}{\sim }f^{\prime }\left( x\right) +%
\frac{\sigma ^{m-l}}{\left( m-l+1\right) !}f^{\left( m-l-1\right) }\left(
x\right) \sum_{k=l}^{m}a_{k}q^{m-l-1}.  \label{2.40}
\end{equation}
{\bf QED}

{\bf Remark}. Formula (\ref{2.35}) defines $U$ as an operator parametrized
by $\sigma $, where $\sigma \in {\Bbb {F}.}$ It may happen that for specific
values of $\sigma $ the operator $U$ could involve less than $p$ points, and
consequently its order would be less than $p$. Once a representation of $U$
as a difference operator is chosen, these points can be easily determined by
solving a linear system of algebraic equations.

\smallskip By way of an example, let us consider the following equation \cite
{KP}: 
\begin{equation}
\Delta ^{3}f\left( x\right) +3\Delta ^{2}f\left( x\right) +\Delta f\left(
x\right) -f\left( x\right) =0.  \label{2.40a}
\end{equation}
If $\Delta =\frac{T-1}{\sigma },\,$for $\sigma =1$ eq. (\ref{2.40a}) becomes 
\[
f\left( x+3\right) -f\left( x+1\right) =0, 
\]
which is of second order, in an appropriate domain.

In the following, $U$ will be assumed to be an operator of order $p$
parametrically depending on $\sigma $, and we shall omit the simple analysis
of the specific cases in which the order could be less than maximal.

\begin{theorem}
If $\Delta \,$is a difference operator of order $p$, then $\widetilde{\Delta 
}=T^{j}\Delta $, $j\in {\Bbb {Z}}$ is a difference operator of the same
order.
\end{theorem}

{\bf Proof.} Let us first prove the result for $j=1$. We have

\[
T\Delta =\frac{1}{\sigma }\sum_{k=l}^{m}a_{k}T^{k+1}=\frac{1}{\sigma }%
\sum_{k=l-1}^{m+1}\widetilde{a}_{k}T^{k},\quad \widetilde{a}_{k}=a_{k-1}.
\]
Hence 
\[
\sum_{k=l+1}^{m+1}\widetilde{a}_{k}=\sum_{k=l}^{m}a_{k}=0
\]
\[
\sum_{k=l+1}^{m+1}k\widetilde{a}_{k}=\sum_{k=l}^{m}\left( k+1\right)
a_{k}=\sum_{k=l}^{m}ka_{k}=1.
\]
Thus, conditions (\ref{2.36}) and (\ref{2.38}) are satisfied for $\widetilde{%
U}$ and that is all that is needed. The proof for $j=-1$ is analogous and
for $j$ arbitrary the result follows by induction. {\bf QED}

Conditions (\ref{2.39}) are not shift invariant. However, once $m\,$and $l\,$%
are chosen equations (\ref{2.39}) can always be imposed. Their solution
depends on $m$ and $l$, not only on the shift invariant difference $m-l$.

\begin{theorem}
The operator $\beta \,$conjugate to the difference operator $\Delta \,$of
eq. (\ref{2.35}) is 
\begin{equation}
\beta =\left( \sum_{k=l}^{m}a_{k}kT^{k}\right) ^{-1}.  \label{2.41}
\end{equation}
\end{theorem}

{\bf Proof.} Using eq. (\ref{2.15}) we have 
\[
\beta =\left( \Delta ^{\prime }\right) ^{-1}=\left[ \Delta ,x\right] ^{-1}. 
\]
Moreover 
\[
\left[ \Delta ,x\right] =\frac{1}{\sigma }\left( \sum_{k=l}^{m}a_{k}\left(
x+k\sigma \right) T^{k}-x\sum_{k=l}^{m}a_{k}T^{k}\right)
=\sum_{k=l}^{m}a_{k}kT^{k} 
\]
and (\ref{2.41}) follows. {\bf QED}

Examples of difference operators and the corresponding operators $\beta \,$%
are $\Delta ^{+}\,$and $\Delta ^{-}\,$of eq. (\ref{2.16}) and (\ref{2.17}).
Both are of order 1. Higher order examples are 
\begin{equation}
\Delta ^{s}=\frac{T-T^{-1}}{2\sigma },\qquad \beta =\left( \frac{T+T^{-1}}{2}%
\right) ^{-1},  \label{2.42}
\end{equation}

\begin{equation}
\Delta ^{\left( III\right) }=-\frac{1}{6\sigma }\left(
T^{2}-6T+3+2T^{-1}\right) ,\qquad \beta =\left( -\frac{T^{2}-3T-T^{-1}}{3}%
\right) ^{-1}  \label{2.43}
\end{equation}
\begin{equation}
\Delta ^{\left( IV\right) }=-\frac{1}{12\sigma }(T^{2}-8T+8T^{-1}-T^{-2}),%
\qquad \beta =\left( -\frac{T^{2}-4T-4T^{-1}+T^{-2}}{6}\right) ^{-1}.
\label{2.44}
\end{equation}
The operators $\Delta ^{s}$, $\Delta ^{\left( III\right) }\,$and $\Delta
^{\left( IV\right) }$ approximate the derivative to order $\sigma ^{2}$, $%
\sigma ^{3}\,$and $\sigma ^{4}\,$respectively.

\begin{theorem}
The expression 
\begin{equation}
P_{n}\left( x\right) \equiv \left( x\beta \right) ^{n}\cdot 1  \label{2.56}
\end{equation}
is a well defined polynomial in $x$ of order $n\,\,$with finite coefficients
depending on a finite number of nonnegative powers of the shifts $\sigma \,$%
for any difference operator $\Delta $. The expression for $P_{n}$ is 
\begin{equation}
P_{n}\left( x\right) =\sum_{k=1}^{n}A_{k}\sigma ^{n-k}x^{k},\qquad A_{n}=1,
\label{2.57}
\end{equation}
where all coefficients $A_{k}\,$are finite and depend only on the
coefficients $a_{k}\,$in the definition of $\Delta $ (see eq. (\ref{2.35})).
In particular, they do not depend on $\sigma $.
\end{theorem}

{\bf Proof.} Let us consider the difference operator $\Delta \,$of eq. (\ref
{2.35}) and define the quantities 
\begin{equation}
\gamma _{j}=\sum_{k=l}^{m}a_{k}k^{j},\quad \gamma _{0}=0,\quad \gamma
_{1}=1,\quad j=0,1,2,....  \label{2.58}
\end{equation}
Let us now prove eq. (\ref{2.57}) by induction. Let $P_{n}\,$be a basic
sequence of polynomials for any $\Delta $, as given by eq. (\ref{2.30}).
Thus we put 
\begin{equation}
P_{n+1}\left( x\right) =\sum_{a=1}^{n+1}B_{a}x^{a},  \label{2.61}
\end{equation}
and must prove that the coefficients $B_{a}$ are finite and depend on $%
\sigma \,$in the proper way (i.e. $B_{a}=\widetilde{B_{a}}\sigma ^{n+1-a}$,
where $\widetilde{B_{a}}$ is finite and does not depend on $\sigma $).

We rewrite eq. (\ref{2.30}) (with $n$ substituted by $n+1$) as 
\begin{equation}
\Delta P_{n+1}=\left( n+1\right) P_{n}.  \label{2.62}
\end{equation}
The left hand side is 
\[
\Delta P_{n+1}=\frac{1}{\sigma }\sum_{b=l}^{m}a_{b}\sum_{a=1}^{n+1}B_{a}%
\left( x+b\sigma \right) ^{a}=\sum_{a=1}^{n+1}B_{a}\sum_{k=0}^{a}\left( 
\begin{array}{l}
a \\ 
k
\end{array}
\right) x^{k}\sigma ^{a-k-1}\gamma _{a-k}
\]
\[
=\sum_{k=0}^{n+1}\sum_{a=k}^{n+1}B_{a}\left( 
\begin{array}{l}
a \\ 
k
\end{array}
\right) x^{k}\sigma ^{a-k-1}\gamma _{a-k}.
\]
Comparing powers on the left and right hand side of eq. (\ref{2.62}), we
obtain a system of linear algebraic equations for the coefficient $B_{k}:$%
\begin{equation}
\sum_{a=k}^{n+1}B_{a}\left( 
\begin{array}{l}
a \\ 
k
\end{array}
\right) \sigma ^{a-k-1}\gamma _{a-k}=\left( n+1\right) A_{k}\sigma ^{n-k}.
\label{2.63}
\end{equation}
The system (\ref{2.63}) has a triangular structure. For $k=n+1$, we get the
identity $0=0$. For $k=n\,$, only one term is present on the left and we get 
$B_{n+1}=1\,$(since we have $\gamma _{0}=0\,$and $A_{n}=1$). The value $%
k=n-1\,$gives 
\[
B_{n}=\sigma \frac{n+1}{n}\left( A_{n-1}-\frac{n}{2}\right) .
\]
In general, the system (\ref{2.63}) implies 
\begin{equation}
B_{n-j}=\sigma ^{j+1}\sum_{k=n-j-1}^{n}\mu _{k}\,A_{k}  \label{2.64}
\end{equation}
where the coefficients $\mu _{k}\,$are easy to calculate, but are cumbersome
(and of little interest), so we do not spell them out. {\bf QED}

Let us present the first few basic polynomials $P_{k}\left( x\right) =\left(
x\beta \right) ^{k}1$ for arbitrary $\Delta $ as given by eq. (\ref{2.35})
with $\gamma _{j}\,$defined in terms of $a_{k}$ by eq. (\ref{2.58}). We
obtain 
\[
P_{0}=\left( x\beta \right) ^{0}\cdot 1=1
\]
\[
P_{1}=\left( x\beta \right) ^{1}\cdot 1=x
\]
\[
P_{2}=\left( x\beta \right) ^{2}\cdot 1=x^{2}-\sigma \gamma _{2}x
\]
\[
P_{3}=\left( x\beta \right) ^{3}\cdot 1=x^{3}-3\sigma \gamma
_{2}x^{2}-\sigma ^{2}\left( \gamma _{3}-3\gamma _{2}^{2}\right) x
\]
\begin{equation}
P_{4}=\left( x\beta \right) ^{4}\cdot 1=x^{4}-6\sigma \gamma
_{2}x^{3}+\sigma ^{2}\left( -4\gamma _{3}+15\gamma _{2}^{2}\right) x^{2}
\label{2.65}
\end{equation}
\[
+\sigma ^{3}\left( -\gamma _{4}+10\gamma _{2}\gamma _{3}-15\gamma
_{2}^{3}\right) x.
\]

For $\sigma \rightarrow 0$, we obviously reobtain the basic series
(sequence) for $\Delta =\partial _{x}$.

For $\Delta ^{+}=\frac{T-1}{\sigma }\,$we have only two values of $a_{j}$,
namely $a_{1}=1$, $a_{0}=-1$, hence $\gamma _{j}=1$, $j=2,3,..$. The
polynomials (\ref{2.65}) in this case reduce to the well--known factorial
powers $P_{n}=x\left( x-\sigma \right) \left( x-2\sigma \right) ...\left(
x-\left( n-1\right) \sigma \right) $.

\subsection{\protect\smallskip Linear Difference Equations and Umbral
Equations}

\smallskip Let us introduce the notation $\widehat{f}=f\left( x\beta \right) 
$, i.e. to each function $f\left( x\right) \in {\frak F}$ we associate an
operator $\widehat{f}\,\in {\frak L}$. We shall consider an operator
equation of the form 
\begin{equation}
\sum_{k=0}^{n}\widehat{A}_{k}U^{k}\widehat{f}=\widehat{g}  \label{2.33}
\end{equation}
where $U$ is a delta operator and $\beta \,$is its conjugate operator
defined in eqs. (\ref{2.14})--(\ref{2.15}). We assume that the operators $%
\widehat{A}_{k}$ and $\widehat{g}$\thinspace can be expanded into formal
power series in $\left( x\beta \right) $.

\begin{definition}
An umbral equation of order $n\,$ is an operator equation of the form (\ref
{2.33}) in which the operators $\widehat{A}_{k}$ and $\widehat{g}$ are
given. The unknown is the operator $\widehat{f}$.
\end{definition}

If $U$ is specified to be $U=\partial _{x}$, then $\beta =1$ and eq. (\ref
{2.33}) reduces to a differential equation of order $n$. If $U$ is a
difference operator, (\ref{2.33})\thinspace is still an operator equation.
Projecting both sides onto a space of functions, i.e. applying them to a
constant, we obtain a difference equation. The order of the difference
equations obtained projecting eq. (\ref{2.33}) may vary depending on the
structure of the operator $\widehat{A}_{k}$ (since it acts on the operator $%
\widehat{f}$) and on the choice of the operator $\Delta $ (and consequently
of $\beta $) in terms of shift operators.

\smallskip Let us first take $U=\partial _{x}$, $\beta =1$ in eq. (\ref{2.33}%
). The obtained linear ODE will have $n$ linearly independent solutions $%
f_{i}\left( x\right) $. We can expand them into formal power series about
any point $x_{0}$, where $x_{0}$ is not a singular point of the equation.
Now let $U=\Delta $ be a difference operator and $\beta $ the corresponding
conjugate operator. Then $f_{i}\left( x\beta \right) 1$ viewed as a formal
power series, will be a solution of the corresponding difference equation.

\begin{definition}
We shall call $\widehat{f}\cdot 1$ an {\bf umbral solution} of the
difference equation 
\begin{equation}
\sum_{k=0}^{n}\widehat{A}_{k}U^{k}\widehat{f}\cdot 1=\widehat{g}\cdot 1
\label{2.34b}
\end{equation}
if the real valued function $f\left( x\right) $ is a solution of the
differential equation 
\begin{equation}
\sum_{k=0}^{n}A_{k}\partial _{x}^{k}f\left( x\right) =g\left( x\right) .
\label{2.34c}
\end{equation}
\end{definition}

\smallskip

Thus each solution of the ODE (\ref{2.34c}) provides a formal power solution
of the difference equation (\ref{2.34b}) (and of the umbral equation (\ref
{2.33})). However, eq. (\ref{2.34b}) and (\ref{2.33}) may have other
solutions. Indeed, for a linear difference equation with constant
coefficients we have the following theorem.

\begin{theorem}
Let $U$ be a difference operator of order $p$ and let us assume that the
operators $\widehat{A}_{k}\,$in eq. (\ref{2.33}) are constant. Equation (\ref
{2.34b}) will then have $np\,$linearly independent solutions, $n\,$of them
umbral ones.
\end{theorem}

{\bf Proof. }Eq. (\ref{2.34b}) in this case is a difference equation
involving $np+1\,$different points. Hence to obtain a solution in a new
point we must specify initial conditions in $np\,$points. This provides $np$
linearly independent solutions, uniquely defined in the lattice points $%
x_{n}=x_{0}+n\sigma $ \cite{KP, Elaydi}. Now, let us consider the continuous
limit of eq. (\ref{2.34b}). It is a linear partial differential equation of
order $n$, possessing analytic solutions which can be expanded around any
nonsingular point. Applying the umbral correspondence to the series
expansion of these solutions, we obtain $n$ solutions of eq. (\ref{2.34b})
which are expressed as formal power series in $\left( x\beta \right) ^{k}$,
and therefore are elements of the algebra ${\frak {F}}$. These are the
umbral solutions admitted by eq. (\ref{2.34b}). The remaining $\left(
n-1\right) p\,$do not belong to ${\frak {F}}$. {\bf QED}

When $\widehat{A}_{k}$ are polynomials in $\left( x\beta \right) $, then
additional shifts may appear in the explicit form of the equations coming
from the umbral equation (\ref{2.33}) via projection and their order may be
different than $np$.

\smallskip As an example, let us consider the ''umbral Airy equation'' 
\begin{equation}
\left[ \Delta ^{2}+a\,x\beta \right] \,\widehat{\Psi }=0,\qquad a=const.
\label{2.80}
\end{equation}
For $\Delta ^{+}=\left( T-1\right) /\sigma $, $\beta =T^{-1}$ and $\stackrel{%
\sim }{\Psi }$ $\left( x\right) =\widehat{\Psi }\cdot 1\,$ we have 
\[
\frac{1}{\sigma ^{2}}\left[ \stackrel{\sim }{\Psi }\left( x+2\sigma \right)
-2\stackrel{\sim }{\Psi }\left( x+\sigma \right) +\stackrel{\sim }{\Psi }%
\left( x\right) \right] +ax\stackrel{\sim }{\Psi }\left( x-\sigma \right) =0.
\]
This is a third order difference equation since it involves the function $%
\stackrel{\sim }{\Psi }\left( x\right) $ at the points $x+2\sigma $, $%
x+\sigma $, $x\,$and $x-\sigma $. For $\Delta =\Delta ^{s}$ eq. (\ref{2.80})
would seem to involve infinitely many points: 
\begin{equation}
\frac{1}{4\sigma ^{2}}\left[ \stackrel{\sim }{\Psi }\left( x+2\sigma \right)
-2\stackrel{\sim }{\Psi }\left( x\right) +\stackrel{\sim }{\Psi }\left(
x-2\sigma \right) \right] +ax\left( \frac{T+T^{-1}}{2}\right) ^{-1}\stackrel{%
\sim }{\Psi }\left( x\right) =0.  \label{2.81}
\end{equation}
However, multiplying eq. (\ref{2.81}) by $\beta ^{-1}\,$we obtain 
\begin{equation}
\frac{1}{4\sigma ^{2}}\left[ \stackrel{\sim }{\Psi }\left( x+3\sigma \right)
-\stackrel{\sim }{\Psi }\left( x+\sigma \right) -\stackrel{\sim }{\Psi }%
\left( x-\sigma \right) +\stackrel{\sim }{\Psi }\left( x-3\sigma \right)
\right] +ax\stackrel{\sim }{\Psi }\left( x\right) =0  \label{2.82}
\end{equation}
This equation is a 6th order difference equation.

As a simple example of umbral and nonumbral solutions, let us consider a
first order homogeneous umbral equation with constant coefficients: 
\begin{equation}
U\widehat{f}=a\widehat{f},\qquad a\neq 0.  \label{2.83}
\end{equation}
For $U=\partial _{x}$, the solution is 
\begin{equation}
f\left( x\right) =Ae^{ax}.  \label{2.84}
\end{equation}
Now, let us consider the first order difference operator $\Delta ^{+}$. Eq. (%
\ref{2.83}) reduces to 
\begin{equation}
f\left( x+\sigma \right) -f\left( x\right) =a\sigma f\left( x\right) .
\label{2.85}
\end{equation}
We look for a solution in the form $f\left( x\right) =\lambda ^{x}\,$and
find 
\begin{equation}
\lambda =\left( 1+a\sigma \right) ^{\frac{1}{\sigma }}.  \label{2.86}
\end{equation}
Thus we obtain a single solution 
\begin{equation}
f_{1}\left( x\right) =A\left( 1+a\sigma \right) ^{\frac{x}{\sigma }}
\label{2.87}
\end{equation}
and of course we have 
\begin{equation}
\lim_{\sigma \rightarrow 0}f\left( x\right) =Ae^{ax}.  \label{2.88}
\end{equation}
The umbral correspondence provides the solution 
\begin{equation}
f_{u}\left( x\right) =Ae^{axT^{-1}}\cdot 1.  \label{2.89}
\end{equation}
Expanding (\ref{2.87}) and (\ref{2.89}) in formal power series in $a$, we
find that the two series coincide, i.e. $f_{1}=f_{u}$.

For comparison, let us consider the second order difference operator $\Delta
^{s}$. Equation (\ref{2.83}) in this case yields 
\begin{equation}
f\left( x+\sigma \right) -f\left( x-\sigma \right) =2\sigma af\left(
x\right) .  \label{2.90}
\end{equation}
Putting $f\left( x\right) =\lambda ^{x}$ we obtain two values of $\lambda $
and the general solution of eq. (\ref{2.90}) in this case is 
\begin{equation}
f=A_{1}\left( \sqrt{1+a^{2}\sigma ^{2}}+a\sigma \right) ^{\frac{x}{\sigma }%
}+A_{2}\left( -1\right) ^{\frac{x}{\sigma }}\left( \sqrt{1+a^{2}\sigma ^{2}}%
-a\sigma \right) ^{\frac{x}{\sigma }}=A_{1}f_{1}+A_{2}f_{2}.  \label{2.91}
\end{equation}
The first solution has $e^{ax\,}\,$as its continuous limit. The second one
does not have a limit for $\sigma \rightarrow 0$. The umbral correspondence
provides the solution 
\begin{equation}
f_{u}\left( x\right) =A\,e^{ax\left( \frac{T+T^{-1}}{2}\right) ^{-1}}\cdot 1
\label{2.92}
\end{equation}
(see eq. (\ref{2.42})). Expanding into formal power series in $a\,$we find $%
f_{u}=f_{1}$ and $f_{2}$ is nonumbral.

The question arises whether an expression of the type 
\begin{equation}
\widehat{f}=e^{ax\beta }  \label{2.93}
\end{equation}
is meaningful, at least in the sense of a formal power series. The problem
is that for a general difference operator $\Delta $, the expression for $%
\beta $, given in eq. (\ref{2.41}), is quite complicated. If we expand $%
\beta \,$into a power series in $T$, it will for $m-l\geq 3\,$involve
infinitely many shifts. Convergence problems may arise. Luckily, it is not
eq. (\ref{2.93}) itself that provides the umbral solution of a difference
equation. Rather, it is the projection of the operator $\widehat{f}\,$onto a
space of functions, or formal power series. The expressions that appear in
the corresponding expansions are $P_{n}\left( x\right) =\left( x\beta
\right) ^{n}\cdot 1$ and these are finite polynomials in $x$, and in the
shifts $\sigma $, with well defined finite coefficients (see Theorem 2.4).
As a matter of fact these are the basic polynomials for the difference
operator $\Delta \,$ defined in eq. (\ref{2.35}).

It follows that if we know a solution of the umbral equation (\ref{2.33})
for $U=\partial _{x}$, and have 
\[
f\left( x\right) =\sum_{n=0}^{\infty }\frac{f^{\left( n\right) }\left(
0\right) }{n!}x^{n},
\]
then for $U=\Delta \,$ as in (\ref{2.35}) the corresponding umbral solution
will be 
\[
\widehat{f}\cdot 1=\sum_{n=0}^{\infty }\frac{f^{\left( n\right) }\left(
0\right) }{n!}P_{n}\left( x\right) .
\]
The matter of convergence is a separate issue.

Umbral calculus and specially the umbral correspondence also provide us with
a powerful tool with which to handle symmetries of linear difference
equations, both ordinary and partial ones. On one hand, we can discretize a
linear differential equation, in particular the linear Schr\"{o}dinger
equation, via the (multidimensional) umbral substitutions 
\[
\partial _{x_{i}}\stackrel{\cal R}{\rightarrow \,}\Delta _{x_{i}},\qquad
x_{i}\stackrel{\cal R}{\rightarrow \,}x_{i}\beta _{x_{i}},\qquad \partial
_{t}\stackrel{\cal R}{\rightarrow \,}\Delta _{t},\qquad t\stackrel{\cal R}{%
\rightarrow \,}t\beta _{t}. 
\]
Lie symmetries, both point and generalized ones, of linear differential
equations can be expressed in terms of commuting operators. Since the umbral
correspondence preserves commutation relations, it will also preserve
symmetries. However, we may have more symmetries than in the continuous case
due to the nonumbral solutions of the determining equations.

\section{\protect\smallskip Discretization of the Time Dependent
Schr\"{o}dinger Equation Preserving all Point Symmetries}

Before considering discrete space--time, let us first give a detailed and
rigorous analysis of the point symmetries of the time--dependent
Schr\"{o}dinger equation in continuous space--time. The results will be
presented in a form well suited for the discretization.

\subsection{Point Symmetries and Commuting Operators in Continuous Space-Time
}

We write the Schr\"{o}dinger equation in ${\Bbb {R}}^{n+1}{\Bbb \,}$as 
\[
L\psi =0, 
\]
\begin{equation}
L=i\partial _{t}-H,\qquad H=-\frac{1}{2}\sum_{k=1}^{n}\frac{\partial ^{2}}{%
\partial x_{k}^{2}}+V\left( \overrightarrow{x},t\right) .  \label{3.1}
\end{equation}

A local Lie point symmetry transformation is generated by a vector field
that we write in evolutionary form \cite{Olver} as 
\begin{equation}
{\bf v}^{E}=Q\partial _{\psi }+Q^{*}\partial _{\psi ^{*}},  \label{3.2}
\end{equation}
\begin{equation}
Q=\eta -\tau \frac{\partial \psi }{\partial t}-\xi _{k}\frac{\partial \psi }{%
\partial x_{k}}.  \label{3.3}
\end{equation}
The functions $\eta ,\tau \,$and $\xi _{k}\,$depend on $t$, $\overrightarrow{%
x}$, $\psi \,$and $\psi ^{*}\,$where the star denotes complex conjugation.
These functions are determined from the requirement 
\begin{equation}
pr^{\left( 2\right) }{\bf v}^{E}\left( L\psi \right) \mid _{L\psi =L^{*}\psi
^{*}=0}=0,\qquad pr^{\left( 2\right) }{\bf v}^{E}\left( L^{*}\psi
^{*}\right) \mid _{L\psi =L^{*}\psi ^{*}=0}=0.  \label{3.4}
\end{equation}

The following theorem will provide a basis for studying the symmetries of a
nonrelativistic quantum system.

\begin{theorem}
All Lie point symmetries of the time--dependent Schr\"{o}dinger equation (%
\ref{3.1}) are generated by evolutionary vector fields of the form (\ref{3.2}%
) with 
\begin{equation}
Q=\chi \left( \overrightarrow{x},t\right) +iX\psi ,  \label{3.5}
\end{equation}
\begin{equation}
X=i\left( \tau \left( t\right) \partial _{t}+\xi _{k}\left( \overrightarrow{x%
},t\right) \partial _{x_{k}}-i\phi \left( \overrightarrow{x},t\right)
\right) ,  \label{3.6}
\end{equation}
\begin{equation}
\xi _{k}\left( \overrightarrow{x},t\right) =\frac{1}{2}x_{k}\tau ^{\prime
}-A_{kl}x_{l}+f_{k}\left( t\right) ,  \label{3.7}
\end{equation}
\begin{equation}
\phi \left( \overrightarrow{x},t\right) =\frac{1}{4}\tau ^{\prime \prime
}r^{2}+x_{k}f_{k}^{\prime }+g\left( t\right) +i\left[ \frac{n}{4}\tau
^{\prime }-B\right] ,  \label{3.8}
\end{equation}
where the prime denotes a derivative. The function $\chi \left( 
\overrightarrow{x},t\right) \,$satisfies the Schr\"{o}dinger equation (\ref
{3.1}), $A_{kl}=-A_{lk}\,$and $B$ are real constants. The real functions $%
\tau \left( t\right) $, $f_{k}\left( t\right) $, $g\left( t\right) $ and the
constants $A_{kl}\,$\thinspace depend on the potential and satisfy the
equation 
\begin{equation}
\tau \left( t\right) V_{t}+\xi _{k}\left( \overrightarrow{x},t\right)
V_{x_{k}}+\tau ^{\prime }V+\frac{1}{4}\tau ^{\prime \prime \prime
}r^{2}+x_{k}f_{k}^{\prime \prime }+g^{\prime }=0,  \label{3.9}
\end{equation}
with $r^{2}=\sum_{k=1}^{n}x_{k}^{2}$. Moreover, the linear operator $X\,$
commutes with $L\,$on the solutions of the Schr\"{o}dinger equation 
\begin{equation}
\left[ L,X\right] \psi \mid _{L\psi =0}=0.  \label{3.10}
\end{equation}
\end{theorem}

{\bf Proof.} Eq. (\ref{3.4}) implies a system of determining equations.
Those among them that come from terms involving derivatives of $\psi $, e.g. 
$\psi _{x}\psi _{xx}\,$, $\psi _{xx}$, $\psi _{x}^{k}$, $k\geq 1\,$do not
depend on the potential $V\left( \overrightarrow{x},t\right) $. From them we
obtain the fact that the corresponding transformations are fiber preserving
and linear (inhomogeneous). That is, $Q$ has the form (\ref{3.3}) with $\tau
\,$and $\xi _{k}\,$ independent of $\psi \,$and $\psi ^{*}$. From the same
equations we find that $\tau $ depends only on $t$ and that $\xi _{k}\left( 
\overrightarrow{x},t\right) $ are linear in $\overrightarrow{x}$. Thus $\xi
_{k}\,\,$and $\phi \,$have the form (\ref{3.7}) and (\ref{3.8}),
respectively.

Once these conditions are satisfied, only one determining equation remains,
namely eq. (\ref{3.9}), involving the potential in a crucial manner.

To prove the commutativity relation (\ref{3.10}) we use the compatibility of
the two flows
\begin{equation}
i\frac{\partial \psi }{\partial t}=H\psi ,\qquad \frac{\partial \psi }{%
\partial \lambda }=Q,  \label{3.10b}
\end{equation}
where $\lambda \,$is a group parameter and $Q$ is the characteristic of the
vector field (see eqs. (\ref{3.5})--(\ref{3.8})). Equating the cross
derivatives $\psi _{t\lambda }=\psi _{\lambda t}\,$and using the equation $%
L\chi \left( \overrightarrow{x},t\right) =0$, we obtain 
\begin{equation}
\left[ H,X\right] \psi =-iX_{t}\psi ,  \label{3.13}
\end{equation}
where $\psi $ is any solution of the Schr\"{o}dinger equation. This is
equivalent to eq. (\ref{3.10}). Simply stated: finding point symmetries of
the Schr\"{o}dinger equation is equivalent to finding linear selfadjoint
operators $X$ commuting with $L\,$on the solution set of $L$. {\bf QED}

{\bf Comments.}

\begin{enumerate}
\item  For any potential $V_{k}\left( \overrightarrow{x},t\right) \,$the
function $\chi \left( \overrightarrow{x},t\right) $, the constant $B$ and a
constant $g=g_{0}$ are solutions of eq. (\ref{3.9}). Hence we always have a
''trivial'' symmetry algebra 
\[
S\left( \chi \right) =\chi \left( \overrightarrow{x},t\right) \partial
_{\psi }+\chi ^{*}\left( \overrightarrow{x},t\right) \partial _{\psi
^{*}},\qquad L\chi =0
\]
\begin{equation}
N=\psi \partial _{\psi }+\psi ^{*}\partial _{\psi ^{*}}  \label{3.11}
\end{equation}
\[
E=i\left( \psi \partial _{\psi }-\psi ^{*}\partial _{\psi ^{*}}\right) ,
\]
due to the linearity of the Schr\"{o}dinger equation.

\item  Each symmetry generator ${\bf v}^{E}\,$provides us with a flow that
is by construction compatible with the time flow (\ref{3.1}), that is, we
can simultaneously solve the equations (\ref{3.10b}). The fixed point $%
\partial \psi /\partial \lambda =0\,$ corresponds to group invariant
solutions.

\item  While the result presented in Theorem (\ref{3.1}) is quite simple and
natural, we have not found it explicitly in the literature, so we have
sketched a proof. For other results on point symmetries of linear
differential equations, see. e.g. \cite{Bluman}, \cite{ML} and \cite{BA}.
\end{enumerate}

Let us consider the implications of Theorem 3.1\ for special cases of the
potential $V\left( \overrightarrow{x},t\right) $. We shall omit the
operators (\ref{3.11}) that are present for any potential $V\left( 
\overrightarrow{x},t\right) $.

Let us first consider the free Schr\"{o}dinger equation (for further
discussions, see also \cite{RiW}). For $V\left( \overrightarrow{x},t\right)
=0$ in ${\Bbb {R}}^{3+1}$ the so called ''Schr\"{o}dinger group'' was first
obtained by Niederer \cite{Nie}. For $n$ arbitrary we obtain its
generalization, i.e. the group $Sch\left( n\right) .$ Its Lie algebra can be
written as 
\[
P_{0}=\partial _{t},\quad D=2t\partial _{t}+x_{k}\partial _{x_{k}}-\frac{1}{2%
}\left( \psi \partial _{\psi }+\psi ^{*}\partial _{\psi ^{*}}\right) ,
\]
\[
C=t^{2}\partial _{t}+tx_{k}\partial _{x_{k}}-\frac{1}{2}t\left( \psi
\partial _{\psi }+\psi ^{*}\partial _{\psi ^{*}}\right) +\frac{in}{4}%
r^{2}\left( \psi \partial _{\psi }-\psi ^{*}\partial _{\psi ^{*}}\right) ,
\]
\begin{equation}
L_{ik}=x_{i}\partial _{x_{k}}-x_{k}\partial _{x_{i}},\qquad P_{k}=\partial
_{x_{k}},  \label{3.14}
\end{equation}
\[
B_{k}=t\partial _{x_{k}}+\frac{i}{2}x_{k}\left( \psi \partial _{\psi }-\psi
^{*}\partial _{\psi ^{*}}\right) ,
\]
\[
E=i\left( \psi \partial _{\psi }-\psi ^{*}\partial _{\psi ^{*}}\right) .
\]
The Levi decomposition \cite{Jacob} of this algebra for $n\geq 3$ is 
\begin{equation}
{\frak {L}}\sim \left[ sl\left( 2,{\Bbb R}\right) \oplus O\left( n\right)
\right] \relax{\supset\kern -13pt +}\text{ }H_{n}  \label{3.15}
\end{equation}
where the radical $H_{n}$ is the $n$--dimensional Heisenberg algebra.
Explicitly we have 
\begin{equation}
sl\left( 2,{\Bbb {R}}\right) \sim \left\{ P_{0},D,C\right\} ,\qquad O\left(
n\right) \sim \left\{ L_{ik}\right\} ,\qquad H_{n}\sim \left\{
P_{k},B_{k},E\right\} .  \label{3.15b}
\end{equation}
We included the central element $E$ explicitly in (\ref{3.14}) since it
appears in the derived algebra of the Schr\"{o}dinger algebra (it is also
present in (\ref{3.11})).

Eq. (\ref{3.9}) implies that for a general time independent potential $%
V\left( \overrightarrow{x}\right) $ we have only one additional symmetry
generator to the set given by eqs. (\ref{3.11}), namely time translations $%
P_{0}=\partial _{t}$.

For a central potential $V=V\left( r\right) $, the additional elements are
time translations and rotations 
\begin{equation}
P_{0}=\partial _{t},\qquad L_{ik}=x_{i}\partial _{x_{k}}-x_{k}\partial
_{x_{i}},\qquad 1\leq i\leq k\leq n.  \label{3.16}
\end{equation}

In the case of a translationally invariant potential $V=V\left( x_{n}\right) 
$, the additional symmetry elements are

\[
P_{0}=\partial _{t},P_{j}=\partial _{x_{j}},B_{j}=t\partial
_{x_{j}}-ix_{j}\left( \psi \partial _{\psi }-\psi ^{*}\partial _{\psi
^{*}}\right) ,\quad 1\leq j\leq n-1,
\]
\begin{equation}
L_{ik}=x_{i}\partial _{x_{k}}-x_{k}\partial _{x_{i}},\qquad 1\leq i\leq
k\leq n-1.  \label{3.17}
\end{equation}

\subsection{Symmetries of the discrete time dependent Schr\"{o}dinger
equation}

\smallskip The umbral correspondence, together with Theorem 3.1 provide the
tools necessary for a symmetry preserving discretization of quantum
mechanics.

Indeed, let us consider a discrete space--time, more precisely an $n+1\,$%
dimensional orthogonal and equally spaced lattice with time step $\sigma
_{t} $ and space steps $\sigma _{k}$, $1\leq k\leq n$. In this space we
write a ''Schr\"{o}dinger difference equation'' 
\begin{equation}
L_{D}\psi =0,\qquad L=i\Delta _{t}-H_{D}  \label{4.1}
\end{equation}
\[
H_{D}=-\frac{1}{2}\sum_{k=1}^{n}\Delta _{x_{k}x_{k}}+V\left( x_{1}\beta
_{1},...,x_{n}\beta _{n},t\beta _{t}\right) 
\]
where we have 
\begin{equation}
\left[ \Delta _{x_{k}},x_{k}\beta _{k}\right] =1,\qquad \left[ \Delta
_{t},t\beta _{t}\right] =1.  \label{4.2}
\end{equation}
Each $\Delta _{x_{k}}$, $\Delta _{t}\,$is some chosen difference operator
and $\beta _{k}$, $\beta _{t}$ are the corresponding conjugate operators
satisfying the Heisenberg commutation relations (\ref{4.2}).

The continuous limit of eq. (\ref{4.1}) is eq. (\ref{3.1}), obtained by
taking $\sigma _{k}\rightarrow 0$, $\sigma _{t}\rightarrow 0$, i.e. 
\begin{equation}
\Delta _{x_{k}x_{k}}\rightarrow \frac{\partial ^{2}}{\partial x_{k}^{2}}%
,\qquad \Delta _{t}\rightarrow \partial _{t},\qquad \beta _{i}\rightarrow
1,\qquad \beta _{t}\rightarrow 1.  \label{4.3}
\end{equation}
Let us assume that in the continuous limit the obtained Schr\"{o}dinger
equation (\ref{3.1}) is invariant under some Lie point symmetry group
generated by some evolutionary vector field (\ref{3.2}). Symmetries of
linear difference equations on fixed lattices can also be expressed in terms
of evolutionary vector fields \cite{LVW}, \cite{LW02}. For eq. (\ref{4.1})
we put 
\begin{equation}
{\bf v}_{D}^{E}=Q_{D}\partial _{\psi }+Q_{D}^{*}\partial _{\psi ^{*}}
\label{4.4}
\end{equation}
\[
Q_{D}=\eta _{D}-\tau _{D}\Delta _{t}\psi -\xi _{k_{D}}\Delta _{x_{k}}\psi 
\]
where $\eta _{D}$, $\tau _{D}$ and $\xi _{k_{D}}$ are functions of $%
x_{i}\beta _{i}$, $t\beta _{t}$, $\psi $ and $\psi ^{*}$. The functions $%
\psi $ and $\psi ^{*}$ are to be evaluated at the points $x_{i}\beta _{i}$, $%
t\beta _{t}$.

The prolongation of the vector field (\ref{4.4}) must act on the dependent
variables $\psi $ and $\psi ^{*}$ and on their discrete derivatives $\Delta
_{t}\psi $, $\Delta _{x_{k}x_{k}}\psi $. As in the continuous case, we
require that an infinitesimal transformation 
\[
\widetilde{x_{k}}\widetilde{\beta _{k}}=x_{k}\beta _{k},\qquad \widetilde{t}%
\widetilde{\beta _{t}}=t\beta _{t} 
\]
\begin{equation}
\widetilde{\psi }\left( \widetilde{x_{k}}\widetilde{\beta _{k}},\widetilde{t}%
\widetilde{\beta _{t}}\right) =\psi \left( x_{k}\beta _{k},t\beta
_{t}\right) +\lambda Q,\qquad \lambda \ll 1  \label{4.5}
\end{equation}
should take a solution $\psi $ into a solution $\widetilde{\psi }\,$of the
same equation (in new variables). First of all, we have 
\begin{equation}
\widetilde{\beta _{k}}=\beta _{k},\qquad \widetilde{\beta _{t}}=\beta _{t}
\label{4.6}
\end{equation}
since $\beta _{k}$ and $\beta _{t}\,$ are expressed in terms of shifts
operators and we are considering equations on a fixed (not transforming)
lattice. Eq. (\ref{4.5}) is an infinitesimal transformation in the
evolutionary formalism, hence only the dependent variables transform. The
transformation of the discrete derivatives is given by 
\[
\Delta _{t}\widetilde{\psi }=\Delta _{t}\psi +\lambda \Delta _{t}Q 
\]
\begin{equation}
\Delta _{x_{k}x_{k}}\widetilde{\psi }=\Delta _{x_{k}x_{k}}\psi +\lambda
\Delta _{x_{k}x_{k}}Q  \label{4.7}
\end{equation}
where $\Delta _{t}$, $\Delta _{x_{k}}$, etc. are discrete total derivatives.
One can of course also introduce discrete partial derivatives \cite{LW02},
but we shall not need them here.

In terms of the vector fields ${\bf v}_{D}^{E}\,$of eq. (\ref{4.4}) the
prolongation of ${\bf v}_{D}^{E}\,\,$is 
\begin{equation}
pr{\bf v}_{D}^{E}=Q_{D}\partial _{\psi }+Q_{D}^{t}\partial _{\Delta _{t}\psi
}+Q_{D}^{x_{k}x_{k}}\partial _{\Delta _{x_{k}x_{k}}\psi }+...+c.c.
\label{4.8}
\end{equation}
where $c.c.\,$denotes the complex conjugate terms and we have 
\begin{equation}
Q_{D}^{t}=\Delta _{t}Q_{D},\qquad Q_{D}^{x_{k}x_{k}}=\Delta
_{x_{k}x_{k}}Q_{D}.  \label{4.9}
\end{equation}
The determining equations for the characteristic $Q_{D}\,$are obtained as in
the continuous case, i.e. from the invariance condition 
\begin{equation}
pr{\bf v}_{D}^{E}\left( L_{D}\psi \right) \mid _{L_{D}\psi =L_{D}^{*}\psi
^{*}=0}=0,\qquad pr{\bf v}_{D}^{E}\left( L_{D}^{*}\psi ^{*}\right) \mid
_{L_{D}\psi =L_{D}^{*}\psi ^{*}=0}=0.  \label{4.10}
\end{equation}
From this we conclude that the following theorem holds.

\begin{theorem}
The discrete time--dependent Schr\"{o}dinger equation (\ref{4.1}) allows a
Lie algebra of ''umbral symmetries'' isomorphic to that of its continuous
limit (\ref{3.1}). This Lie algebra is realized by vector fields (\ref{4.4})
with 
\begin{equation}
Q_{D}=\chi \left( x_{k}\beta _{k},t\beta _{t}\right) +iX_{D}\psi 
\label{4.11}
\end{equation}
\begin{equation}
X_{D}=i\left[ \tau \left( t\beta _{t}\right) \Delta _{t}+\sum_{k}\xi
_{k}\Delta _{x_{k}}-i\phi \right]   \label{4.12}
\end{equation}
\begin{equation}
\xi _{k}=\frac{1}{2}x_{k}\beta _{k}\Delta _{t}\tau
-\sum_{l=1}^{n}A_{kl}x_{l}\beta _{l}+f_{k}\left( t\beta _{t}\right) 
\label{4.13}
\end{equation}
\begin{equation}
\phi =\left[ \frac{1}{4}\Delta _{tt}\tau \sum_{k=1}^{n}\left( x_{k}\beta
_{k}\right) ^{2}+\sum_{k=1}^{n}x_{k}\beta _{k}\Delta _{t}f_{k}+g\left(
t\beta _{t}\right) \right] +i\left[ \frac{n}{4}\left( \Delta _{t}\tau
\right) -B\right]   \label{4.14}
\end{equation}
The function $\chi \,$satisfies the discrete Schr\"{o}dinger equation (\ref
{4.1}), $A_{kl}=-A_{lk}$ and $B$ are real constants. The real functions $%
\tau $, $f_{k}$ and $g\,$all depend only on $t\beta _{t}\,$and the potential 
$V\left( x_{k}\beta _{k},t\beta _{t}\right) \,$satisfies: 
\begin{equation}
\tau \Delta _{t}V+\sum_{k=1}^{n}\xi _{k}\Delta _{x_{k}}V+\left( \Delta
_{t}\tau \right) V+\frac{1}{4}\left( \Delta _{ttt}\tau \right)
\sum_{k=1}^{n}\left( x_{k}\beta _{k}\right) ^{2}+\sum_{k=1}^{n}x_{k}\beta
_{k}\Delta _{tt}f_{k}+\Delta _{t}g=0.  \label{4.15}
\end{equation}
Finally, the difference operator $X_{D}\,$commutes with $L_{D}\,$on the
solutions of the discrete Schr\"{o}dinger equation (\ref{4.1}): 
\begin{equation}
\left[ L_{D},X_{D}\right] \psi \mid _{L_{D}\psi =0}\quad =0.  \label{4.16}
\end{equation}
\end{theorem}

{\bf Proof. }The proof of Theorem 3.2 is quite analogous to that of Theorem
3.1 in the continuous case. To see the similarities and differences, let us
restrict ourselves to the case $n=2$.

The invariance condition (\ref{4.10}) implies the following determining
equations 
\begin{equation}
\xi _{k,\psi }=\xi _{k,\psi ^{*}}=\tau _{\psi }=\tau _{\psi ^{*}}=0,\quad
\Delta _{x_{i}}\tau =0,\quad \phi _{\psi ^{*}}=0,\quad \phi _{\psi \psi }=0,
\label{4.17}
\end{equation}
\begin{equation}
\Delta _{x_{1}}\xi _{2}+\Delta _{x_{2}}\xi _{1}=0,  \label{4.18}
\end{equation}
\begin{equation}
\Delta _{t}\tau -2\Delta _{x_{1}}\xi _{1}=0,\qquad \Delta _{t}\tau -2\Delta
_{x_{2}}\xi _{2}=0,  \label{4.19}
\end{equation}
\[
2i\Delta _{t}\xi _{1}+2\Delta _{x_{1}}\phi _{1\psi }+\Delta _{x_{1}x_{1}}\xi
_{1}+\Delta _{x_{2}x_{2}}\xi _{1}=0,
\]
\begin{equation}
2i\Delta _{t}\xi _{2}+2\Delta _{x_{2}}\phi _{1\psi }+\Delta _{x_{1}x_{1}}\xi
_{2}+\Delta _{x_{2}x_{2}}\xi _{2}=0,  \label{4.20}
\end{equation}
\[
2\psi \left\{ \tau \Delta _{t}V+\xi _{1}\Delta _{x_{1}}V+\xi _{2}\Delta
_{x_{2}}V+V\Delta _{t}\tau +V\phi _{\psi }\right\} 
\]
\begin{equation}
-2V\phi +2i\Delta _{t}\phi +\Delta _{x_{1}x_{1}}\phi +\Delta
_{x_{2}x_{2}}\phi =0.  \label{4.21}
\end{equation}

It is now obvious that eq. (\ref{4.12}), (\ref{4.13}) and (\ref{4.14}) (for $%
n=2$) provide a solution to eq. (\ref{4.17}),..., (\ref{4.20}) and that (\ref
{4.21}) reduces to eq. (\ref{4.15}), once (\ref{4.17}),..., (\ref{4.20}) are
solved. Eq. (\ref{4.16}) then follows in exactly the same manner as in the
continuous case. {\bf QED}.

\begin{description}
\item  {\bf Comment.}
\end{description}

There is an important difference between the continuous and the discrete
case. In Theorem 3.1 we presented the most general solution of the
determining equations. In Theorem 3.2 we presented a solution and added the
requirement that the solution should have the correct continuous limit. Take
for instance the function $\tau $. Equations (\ref{4.17}) for any $\Delta
_{t}$, $\Delta _{x}\,$allow the solution $\tau \left( t,\beta _{t}\right) $,
i.e. an arbitrary function of time $t$ and the ''shift'' operator $\beta
_{t} $ independently. For first order operators $\Delta ^{\pm }\,$(see
Section 2), the function $\tau $ will depend only on $\left( t\beta
_{t}\right) $. This follows from the determining equations, and agrees with
the result of umbral correspondence. Moreover, the result will have the
correct continuous limit. Thus the discretization and the continuous
equation have isomorphic symmetry algebras. For other choices of the
discrete derivatives we may get more general solutions. For instance, let us
consider the case of a ''symmetric'' derivative: 
\begin{equation}
\Delta _{x}^{s}\tau =\frac{T_{x}-T_{x}^{-1}}{2\sigma }\tau =\frac{\tau
\left( x+\sigma \right) -\tau \left( x-\sigma \right) }{2\sigma }=0.
\label{4.22}
\end{equation}
Eq. (\ref{4.22}) has the general solution 
\begin{equation}
\tau =\tau _{0}\left( t\right) +\tau _{1}\left( t\right) e^{\frac{i\pi x}{%
\sigma }},  \label{4.23}
\end{equation}
\begin{equation}
x=x_{n}=x_{0}+n\sigma .  \label{4.24}
\end{equation}
We see that eq. (\ref{4.22}) actually allows an $x$--dependence in $\tau $.
However, the second term in (\ref{4.24}) does not have a continuous limit
(for $\sigma =x_{n+1}-x_{n}\rightarrow 0$).

\subsection{Examples}

As in the continuous case, for any discrete potential we have the
''trivial'' symmetries (\ref{3.11}) (with $\overrightarrow{x}$, $t$ replaced
by $x_{k}\beta _{k}$, $t\beta _{t}$).

Let us consider the case $V=0$, i.e. a free quantum particle in discrete
space time of dimension $n+1$. Operators commuting with the operator $L$ of
eq. (\ref{4.1}) are obtained from (\ref{3.14}) by the umbral correspondence.
We obtain the ''discrete'' Schr\"{o}dinger algebra 
\[
P_{0}=\Delta _{t},\qquad D=2\left( t\beta _{t}\right) \Delta
_{t}+\sum_{k=1}^{n}\left( x_{k}\beta _{k}\right) \Delta _{x_{k}}-\frac{1}{2}%
\left( \psi \partial _{\psi }+\psi ^{*}\partial _{\psi ^{*}}\right) 
\]
\[
C=\left( t\beta _{t}\right) ^{2}\Delta _{t}+\sum_{k=1}^{n}\left( t\beta
_{t}\right) \left( x_{k}\beta _{k}\right) \Delta _{x_{k}}-\frac{1}{2}\left(
t\beta _{t}\right) \left( \psi \partial _{\psi }+\psi ^{*}\partial _{\psi
^{*}}\right) +
\]
\[
\frac{in}{4}\sum_{k=1}^{n}\left( x_{k}\beta _{k}\right) ^{2}\left( \psi
\partial _{\psi }-\psi ^{*}\partial _{\psi ^{*}}\right) 
\]
\[
L_{ik}=\left( x_{i}\beta _{i}\right) \Delta _{x_{k}}-\left( x_{k}\beta
_{k}\right) \Delta _{x_{i}},\qquad P_{k}=\Delta _{x_{k}},
\]
\[
B_{k}=\left( t\beta _{t}\right) \Delta _{x_{k}}+\frac{i}{2}\left( x_{k}\beta
_{k}\right) \left( \psi \partial _{\psi }-\psi ^{*}\partial _{\psi
^{*}}\right) 
\]
\begin{equation}
E=i\left( \psi \partial _{\psi }-\psi ^{*}\partial _{\psi ^{*}}\right) .
\label{4.26}
\end{equation}

For a general time independent potential $V\left( x_{i}\beta _{i}\right) $
we have only one additional (to (\ref{3.11})) symmetry generator, namely
time translations $P_{0}=\Delta _{t}$.

For a time independent central potential $V=V\left( \sum_{i}\left(
x_{i}\beta _{i}\right) ^{2}\right) $ the additional symmetries are expressed
by the operators 
\begin{equation}
P_{0}=\Delta _{t},\qquad L_{ik}=x_{i}\beta _{i}\Delta _{x_{k}}-x_{k}\beta
_{k}\Delta _{x_{i}},\qquad 1\leq i\leq k\leq n.  \label{4.27}
\end{equation}

For a translationally invariant potential $V=V\left( x_{n}\beta _{n}\right) $
the additional symmetry operators are

\[
P_{0}=\Delta _{t},
\]
\[
P_{j}=\Delta _{x_{j}},\quad B_{j}=t\beta _{t}\Delta _{x_{j}}-ix_{j}\beta
_{j}\left( \psi \Delta _{\psi }-\psi ^{*}\Delta _{\psi ^{*}}\right) ,\quad
1\leq j\leq n-1
\]
\begin{equation}
L_{ik}=x_{i}\beta _{i}\Delta _{x_{k}}-x_{k}\beta _{k}\Delta _{x_{i}},\qquad
1\leq i\leq k\leq n-1.  \label{4.28}
\end{equation}

\section{Discrete Superintegrable Systems}

The umbral calculus provides a systematic method for transfering results
from standard quantum mechanics to quantum mechanics in a discrete
space--time. This is particularly simple if the results are formulated in
terms of commuting differential operators. It has been shown elsewhere \cite
{STW} that there is a direct relation between generalized symmetries in
quantum mechanics and higher order differential operators, commuting with
the Hamiltonian. Here we shall briefly sum up the results and then adapt
them to the discrete case.

\subsection{Generalized Symmetries in Quantum Mechanics}

Let us consider the stationary Schr\"{o}dinger equation in real
two--dimensional Euclidean space 
\begin{equation}
H\psi =E\psi ,\qquad H=-\frac{1}{2}\Delta +V\left( x,y\right) ,  \label{5.1}
\end{equation}
and look for second order generalized symmetries in their evolutionary form $%
{\bf v}^{E}$ (\ref{3.2}) with characteristic $Q\,$ satisfying 
\begin{equation}
Q=Q\left( x,y,\psi ,\psi _{x},\psi _{y},\psi _{xx},\psi _{xy},\psi
_{yy}\right) .  \label{5.2}
\end{equation}
We require that the second prolongation of the vector field ${\bf v}^{E}$
should annihilate eq. (\ref{5.1}) on its solution space, i.e. 
\begin{equation}
pr^{\left( 2\right) }{\bf v}^{E}\left( H-E\right) \psi \mid \Sb H\psi =E\psi 
\\ H\psi ^{*}=E\psi ^{*}  \endSb =0,\qquad pr^{\left( 2\right) }{\bf v}%
^{E}\left( H-E\right) \psi \mid \Sb H\psi =E\psi  \\ H\psi ^{*}=E\psi ^{*} 
\endSb =0.  \label{5.3}
\end{equation}
If we also require that $Q$ be energy independent, we obtain the following
result.

\begin{theorem}
The characteristic $Q\,$ of the evolutionary vector field ${\bf v}%
^{E}=Q\partial _{\psi }+Q^{*}\partial _{\psi ^{*}}$, corresponding to a
second order generalized symmetry of the Schr\"{o}dinger equation (\ref{5.1}%
) has the form 
\begin{equation}
Q=X\psi +\chi \left( x,y\right)   \label{5.4}
\end{equation}
\begin{eqnarray}
X &=&aL_{3}^{2}+b\left( L_{3}P_{1}+P_{1}L_{3}\right) +c\left(
L_{3}P_{2}+P_{2}L_{3}\right) +d\left( P_{1}^{2}-P_{2}^{2}\right) 
\label{5.5} \\
&&+2eP_{1}P_{2}+\alpha L_{3}+\beta P_{1}+\gamma P_{2}+\phi (x,y).  \nonumber
\end{eqnarray}
The function $\chi \left( x,y\right) $ satisfies the Schr\"{o}dinger
equation (\ref{5.1}). The operator $X$ commutes with the Hamiltonian $H$%
\begin{equation}
\left[ H,X\right] =0.  \label{5.6}
\end{equation}
The quantities $a,...,e,\alpha ,\beta ,\gamma $ are constants and 
\begin{equation}
P_{1}=\partial _{x},\quad P_{2}=\partial _{y},\quad L_{3}=y\partial
_{x}-x\partial _{y}  \label{5.6b}
\end{equation}
are generators of the Euclidean group $E_{2}$.
\end{theorem}

For a proof of Theorem 4.1, see Ref. \cite{STW}.

The commutativity relation (\ref{5.6}) is equivalent to the following linear
partial differential equations satisfied by the potential $V\left(
x,y\right) \,$and the function $\phi \left( x,y\right) $%
\begin{equation}
\left[ \alpha \left( y\partial _{x}-x\partial _{y}\right) +\beta \partial
_{x}+\gamma \partial _{y}\right] V\left( x,y\right) =0,  \label{5.7}
\end{equation}
\[
(-axy-bx+cy+e)(V_{xx}-V_{yy})+[a(x^{2}-y^{2})-2by-2cx-2d]V_{xy} 
\]
\begin{equation}
\,-3(ay+b)V_{x}+3(ax-c)V_{y}=0,  \label{5.8}
\end{equation}
\begin{equation}
\phi _{x}=-2(ay^{2}+2by+d)V_{x}+2(axy+bx-cy-e)V_{y},  \label{5.8b}
\end{equation}
\begin{equation}
\phi _{y}=2(axy+bx-cy-e)V_{x}+2(-ax^{2}+2cx+d)V_{y}.  \label{5.9}
\end{equation}

Here eq. (\ref{5.8}) is the compatibility condition for the two equations (%
\ref{5.8b}) and (\ref{5.9}). Eq. (\ref{5.7}) is easily solved.

For $\alpha \neq 0$ we can translate $x$ and $y\,$ to transform $\beta
\rightarrow 0$, $\gamma \rightarrow 0$. Then the potential is rotationally
invariant: $V=V\left( r\right) $.

For $\alpha =0,$ $\beta ^{2}+\gamma ^{2}\neq 0$ we can rotate to obtain $%
\beta \rightarrow 0$. Then the potential is translationally invariant: $%
V=V\left( x\right) $.

To avoid the geometric symmetries (\ref{5.6b}) we solve eq. (\ref{5.7})
trivially by imposing $\alpha =\beta =\gamma =0$. We then simplify the
second order operator $X\,$of eq. (\ref{5.5}) by rotations, translations and
linear combinations with the Hamiltonian $H$.

These transformations leave two expressions in the space of the coefficients 
$a,...,e$ invariant, namely 
\begin{equation}
I_{1}=a,\qquad I_{2}=\left[ \left( 2ad-b^{2}+c^{2}\right) ^{2}+4\left(
ae-bc\right) ^{2}\right] .  \label{5.10}
\end{equation}
In the nongeneric case when $I_{1}=I_{2}=0$, a third invariant exists,
namely 
\begin{equation}
I_{3}=d^{2}+e^{2}.  \label{5.10b}
\end{equation}
Using these invariants, one obtains 4 equivalence classes of operators $X$
and correspondingly, four classes of potentials allowing for the existence
of an operator $X$, commuting with the Hamiltonian \cite{WSUF}, \cite{STW}.
The existence of one second order operator $X\,$, satisfying (\ref{5.6})
makes the system integrable. Moreover, the corresponding Schr\"{o}dinger
equation will allow separation of variables in Cartesian, polar, parabolic
or elliptic coordinates, which in the separable system depends on the values
of the invariants (\ref{5.10}) and (\ref{5.10b}).

We are interested in the case of superintegrable Hamiltonians, when two
operators $X_{1}$ and $X_{2}$ exist, satisfying 
\begin{equation}
\left[ H,X_{1}\right] =\left[ H,X_{2}\right] =0,\qquad \left[
X_{1},X_{2}\right] \neq 0.  \label{5.11}
\end{equation}
Four classes of such potentials exist, each allowing the separation of
variables in at least two coordinate systems. The Hamiltonians and
corresponding integrals of motion are:

\begin{enumerate}
\item  
\begin{equation}
H_{I}=-\frac{1}{2}\left( \partial _{x}^{2}+\partial _{y}^{2}\right) +\frac{%
\omega }{2}^{2}(x^{2}+y^{2})+\frac{a}{2x^{2}}+\frac{b}{2y^{2}},  \label{5.12}
\end{equation}
\[
\widehat{X}_{1}=P_{1}^{2}-P_{2}^{2}-\left[ \omega ^{2}(x^{2}-y^{2})+\frac{a}{%
x^{2}}-\frac{b}{y^{2}}\right] ,
\]
\[
\widehat{X}_{2}=L_{3}^{2}-\left( \frac{a}{\cos ^{2}\phi }+\frac{b}{\sin
^{2}\phi }\right) ,
\]
\[
x=r\cos \phi ,\quad y=r\sin \phi .
\]

\item  
\begin{equation}
H_{II}=-\frac{1}{2}\left( \partial _{x}^{2}+\partial _{y}^{2}\right) +\omega
^{2}(2x^{2}+\frac{y^{2}}{2})+\frac{a}{2y^{2}}+bx  \label{5.13}
\end{equation}
\[
\widehat{X}_{1}=P_{1}^{2}-P_{2}^{2}-\left[ \omega ^{2}(4x^{2}-y^{2})+bx-%
\frac{a}{y^{2}}\right] 
\]
\[
\widehat{X}_{2}=L_{3}P_{2}+P_{2}L_{3}-2\,\omega ^{2}\,x\,y^{2}+\frac{2\,ax}{%
y^{2}}-by^{2}
\]
The remaining two systems are best written in parabolic coordinates

\begin{equation}
x=\frac{1}{2}\left( \xi ^{2}-\eta ^{2}\right) ,\quad y=\xi \eta .
\label{5.14b}
\end{equation}

\item  
\begin{equation}
H_{III}=-\frac{1}{2}\frac{1}{\xi ^{2}+\eta ^{2}}\left( \partial _{\xi
}^{2}+\partial _{\eta }^{2}\right) +\frac{1}{\xi ^{2}+\eta ^{2}}\left( 2a+%
\frac{b}{\xi ^{2}}+\frac{c}{\eta ^{2}}\right)   \label{5.14}
\end{equation}
\[
X_{1}=L_{3}^{2}-2\left( \xi ^{2}+\eta ^{2}\right) \left( \frac{b}{\xi ^{2}}+%
\frac{c}{\eta ^{2}}\right) 
\]
\[
X_{2}=L_{3}P_{2}+P_{2}L_{3}+\frac{2}{\xi ^{2}+\eta ^{2}}\left( a\left( \xi
^{2}-\eta ^{2}\right) -b\frac{\eta ^{2}}{\xi ^{2}}+c\frac{\xi ^{2}}{\eta ^{2}%
}\right) .
\]

(For $b=c=0$, $a\neq 0$ this is the Coulomb atom). The system allows
separation of variables in polar and parabolic coordinates (and also in
elliptic coordinates).

\item  
\begin{equation}
H_{IV}=-\frac{1}{2}\frac{1}{\xi ^{2}+\eta ^{2}}\left( \partial _{\xi
}^{2}+\partial _{\eta }^{2}\right) +\frac{2a+b\xi +c\eta }{\xi ^{2}+\eta ^{2}%
}  \label{5.15}
\end{equation}
\[
\widehat{X}_{1}=L_{3}P_{1}+P_{1}L_{3}+\frac{b\eta \left( \eta ^{2}-\xi
^{2}\right) +c\xi \left( \xi ^{2}-\eta ^{2}\right) -4a\eta \xi }{\left( \xi
^{2}+\eta ^{2}\right) }
\]
\[
\widehat{X}_{2}=L_{3}P_{2}+P_{2}L_{3}+2\frac{a\,(\xi ^{2}-\eta ^{2})+\eta
\xi \,(c\xi -b\eta )}{(\xi ^{2}+\eta ^{2})}
\]

The equation separates in two mutually orthogonal parabolic coordinate
systems, namely (\ref{5.14b}) and a similar system with $x$ and $y$
interchanged.

For $a\neq 0$, $b=c=0\,$ we again obtain the Coulomb atom.
\end{enumerate}

We shall call the systems $H_{I}$ and $H_{II}$ the {\it generalized isotropic%
} and {\it generalized nonisotropic harmonic oscillators}, respectively.
Similarly, $H_{III}$ and $H_{IV}\,$can both be called {\it generalized
Coulomb systems}.

The Schr\"{o}dinger equations for $H_{III}$ and $H_{IV}$ can be rewritten as

\begin{equation}
\left\{ -\frac{1}{2}\left( \partial _{\xi }^{2}+\partial _{\eta }^{2}\right)
-E\left( \xi ^{2}+\eta ^{2}\right) +\frac{b}{2\xi ^{2}}+\frac{c}{2\eta ^{2}}%
\right\} \psi =-a\psi ,  \label{5.16}
\end{equation}

\begin{equation}
\left\{ -\frac{1}{2}\left( \partial _{\xi }^{2}+\partial _{\eta }^{2}\right)
-E\left[ \left( \xi -\frac{b}{2E}\right) ^{2}+\left( \eta -\frac{c}{2E}%
\right) ^{2}\right] \right\} \psi =  \label{5.18}
\end{equation}
\[
=\left( -2a-\frac{b^{2}+c^{2}}{4E^{2}}\right) \psi , 
\]
respectively. Thus, the system $H_{III}$ is reduced to $H_{I}$ with the
energy ($-E$) and coupling constant $\omega ^{2}$ interchanged. The system $%
H_{IV}$ is reduced to a ''shifted'' harmonic oscillator. This interchange of
the energy and a coupling constant has been called ''metamorphosis of the
coupling constant'' \cite{HGDR}.

\subsection{Discrete Generalized Harmonic Oscillators}

The umbral correspondence immediately provides us with discrete versions of
these systems.

Let us first consider the potential $V_{I}\,$of eq. (\ref{5.12}). The
discrete version of this system is:

\begin{equation}
H_{I}^{D}=-\frac{1}{2}\left( \Delta _{x}^{2}+\Delta _{y}^{2}\right) +\frac{%
\omega ^{2}}{2}\left[ \left( x\beta _{x}\right) ^{2}+\left( y\beta
_{y}\right) ^{2}\right] +\frac{a}{2}\left( x\beta _{x}\right) ^{-2}+\frac{b}{%
2}\left( y\beta _{y}\right) ^{-2}  \label{5.27}
\end{equation}
with the integrals of motion 
\begin{equation}
X_{1}=\left[ -\frac{1}{2}\Delta _{x}^{2}+\omega ^{2}\left( x\beta
_{x}\right) ^{2}+a\left( x\beta _{x}\right) ^{-2}\right] -\left[ -\frac{1}{2}%
\Delta _{y}^{2}+\omega ^{2}\left( y\beta _{y}\right) ^{2}+b\left( y\beta
_{y}\right) ^{-2}\right]  \label{5.28}
\end{equation}
and 
\[
X_{2}=\left( x\beta _{x}\Delta _{y}-y\beta _{y}\Delta _{x}\right) ^{2} 
\]
\begin{equation}
-\left[ a\left( 1+\left( x\beta _{x}\right) ^{-2}\left( y\beta _{y}\right)
^{2}\right) +b\left( 1+\left( x\beta _{x}\right) ^{2}\left( y\beta
_{y}\right) ^{-2}\right) \right] .  \label{5.29}
\end{equation}

Similarly, the discrete version of the system with potential $V_{II}\,$is 
\begin{equation}
H_{II}^{D}=-\frac{1}{2}\left( \Delta _{x}^{2}+\Delta _{y}^{2}\right) +\omega
^{2}\left[ 2\left( x\beta _{x}\right) ^{2}+\frac{1}{2}\left( y\beta
_{y}\right) ^{2}\right] +\frac{a}{2}\left( y\beta _{y}\right) ^{-2}+bx\beta
_{x}.  \label{5.30}
\end{equation}
The second order operators commuting with the Hamiltonian (\ref{5.30}) are

\begin{equation}
X_{1}=\Delta _{x}^{2}-\Delta _{y}^{2}-\left[ \omega ^{2}\left( 4\left(
x\beta _{x}\right) ^{2}-\left( y\beta _{y}\right) ^{2}\right) +b\left(
x\beta _{x}\right) -a\left( y\beta _{y}\right) ^{-2}\right] ,  \label{5.31}
\end{equation}
and 
\[
X_{2}=\left[ \left( y\beta _{y}\right) \Delta _{x}-\left( x\beta _{x}\right)
\Delta _{y}\right] \Delta _{y}+\Delta _{y}\left[ \left( y\beta _{y}\right)
\Delta _{x}-\left( x\beta _{x}\right) \Delta _{y}\right] -2\omega ^{2}\left(
x\beta _{x}\right) \left( y\beta _{y}\right) ^{2} 
\]

\begin{equation}
+2a\left( x\beta _{x}\right) \left( y\beta _{y}\right) ^{-2}-b\left( y\beta
_{y}\right) ^{2}.  \label{5.32}
\end{equation}

\subsection{Discrete generalized Coulomb potentials}

To discretize the systems $H_{III}$ and $H_{IV}\,$we again use the umbral
correspondence, this time using parabolic coordinates. Thus, we replace

\begin{equation}
\partial _{\xi }\rightarrow \Delta _{\xi },\quad \partial _{\eta
}\rightarrow \Delta _{\eta },\quad \xi \rightarrow \xi \beta _{\xi },\quad
\eta \rightarrow \eta \beta _{\eta }.  \label{5.33}
\end{equation}

With these replacements it is a simple matter to write the discrete versions
of the systems corresponding to the potentials $V_{III}\,$and $V_{IV}$.
Indeed, we have 
\begin{equation}
H_{III}=-\frac{1}{2}\left[ \left( \xi \beta _{\xi }\right) ^{2}+\left( \eta
\beta _{\eta }\right) ^{2}\right] ^{-1}\left[ \Delta _{\xi }^{2}+\Delta
_{\eta }^{2}-4a-2b\left( \xi \beta _{\xi }\right) ^{-2}-2c\left( \eta \beta
_{\eta }\right) ^{-2}\right]   \label{5.34}
\end{equation}
\begin{equation}
X_{1}=\left[ \left( \xi \beta _{\xi }\right) \Delta _{\eta }-\left( \eta
\beta _{\eta }\right) \Delta _{\xi }\right] ^{2}-2\left[ \left( \xi \beta
_{\xi }\right) ^{2}+\left( \eta \beta _{\eta }\right) ^{2}\right] \left[
b\left( \xi \beta _{\xi }\right) ^{-2}+c\left( \eta \beta _{\eta }\right)
^{-2}\right]   \label{5.35}
\end{equation}
\[
X_{2}=\left[ \left( \xi \beta _{\xi }\right) ^{2}+\left( \eta \beta _{\eta
}\right) ^{2}\right] ^{-1}\left\{ \left( \eta \beta _{\eta }\right)
^{2}\Delta _{\xi }^{2}-\left( \xi \beta _{\xi }\right) ^{2}\Delta _{\eta
}^{2}+2a\left[ \left( \xi \beta _{\xi }\right) ^{2}-\left( \eta \beta _{\eta
}\right) ^{2}\right] \right. 
\]
\begin{equation}
\left. -2b\left( \eta \beta _{\eta }\right) ^{2}\left( \xi \beta _{\xi
}\right) ^{-2}+2c\left( \xi \beta _{\xi }\right) ^{2}\left( \eta \beta
_{\eta }\right) ^{-2}\right\}   \label{5.36}
\end{equation}

and 
\begin{equation}
H_{IV}=-\frac{1}{2}\left[ \left( \xi \beta _{\xi }\right) ^{2}+\left( \eta
\beta _{\eta }\right) ^{2}\right] ^{-1}\left[ \Delta _{\xi }^{2}+\Delta
_{\eta }^{2}-4a-2b\left( \xi \beta _{\xi }\right) -2c\left( \eta \beta
_{\eta }\right) \right]   \label{5.37}
\end{equation}
\[
X_{1}=\left[ \left( \xi \beta _{\xi }\right) ^{2}+\left( \eta \beta _{\eta
}\right) ^{2}\right] ^{-1}\left\{ \left( \xi \beta _{\xi }\right) \left(
\eta \beta _{\eta }\right) \left( \Delta _{\xi }^{2}+\Delta _{\eta
}^{2}\right) +\right. 
\]
\begin{equation}
\left. \left[ -b\left( \eta \beta _{\eta }\right) +c\left( \xi \beta _{\xi
}\right) \right] \left[ \left( \xi \beta _{\xi }\right) ^{2}-\left( \eta
\beta _{\eta }\right) ^{2}\right] -4a\left( \xi \beta _{\xi }\right) \left(
\eta \beta _{\eta }\right) \right\} -\Delta _{\xi \eta }^{2}.  \label{5.38}
\end{equation}
\[
X_{2}=\frac{1}{2}\left[ \left( \xi \beta _{\xi }\right) ^{2}+\left( \eta
\beta _{\eta }\right) ^{2}\right] ^{-1}\left\{ \left( \eta \beta _{\eta
}\right) ^{2}\Delta _{\xi }^{2}-\left( \xi \beta _{\xi }\right) ^{2}\Delta
_{\eta }^{2}+2a\left[ \left( \xi \beta _{\xi }\right) ^{2}-\left( \eta \beta
_{\eta }\right) ^{2}\right] \right. 
\]
\begin{equation}
\left. +2\left( \xi \beta _{\xi }\right) \left( \eta \beta _{\eta }\right)
\left[ c\left( \xi \beta _{\xi }\right) -b\left( \eta \beta _{\eta }\right)
\right] \right\}   \label{5.39}
\end{equation}

\section{Exact Solvability and Spectral Properties of Discrete
Superintegrable Systems}

We have shown that certain important properties of the Schr\"{o}dinger
equation, such as point and generalized symmetries, and hence also
integrability, are preserved when we pass from continuous to discrete
space--time via an umbral correspondence.

Another important property of some quantum systems is their ''exact
solvability''. This means that their Hamiltonian can be transformed into a
block diagonal form with finite--dimensional blocks. In other words, their
complete energy spectrum can be calculated algebraically. In more
mathematical terms, we give the following definition.

\begin{definition}
A quantum mechanical system with Hamiltonian $H$ is called exactly solvable
if its Hilbert space $S\,$of bound states consists of a flag of finite
dimensional subspaces 
\begin{equation}
S_{0}\subset S_{1}\subset S_{2}\subset ...S_{n}\subset ...  \label{6.1}
\end{equation}
preserved by the Hamiltonian 
\begin{equation}
HS_{i}\subseteq S_{i}.  \label{6.2}
\end{equation}
\end{definition}

All known exactly solvable systems also have the following properties.

\begin{enumerate}
\item  In appropriate coordinates and in an appropriate gauge, the bound
state wave functions $\Psi _{N}$\smallskip $\left( \overrightarrow{x}\right) 
$ are polynomials: 
\begin{equation}
\Psi _{N}\smallskip \left( \overrightarrow{x}\right) =g\smallskip \left( 
\overrightarrow{x}\right) P_{N}\smallskip \left( \overrightarrow{s}\right)
,\qquad s_{i}=s_{i}\smallskip \left( \overrightarrow{x}\right) .  \label{6.3}
\end{equation}
The gauge factor $g\smallskip \left( \overrightarrow{x}\right) $ is a priori
defined and can be energy dependent. The function $P_{N}\smallskip \left( 
\overrightarrow{s}\right) $ are polynomials of order $N$ in the variables $%
s_{i}$. The integer $N\,$labels the subspaces $S_{i}\,$in the flag.

\item  In the same gauge $g$ and same variables $s_{i}$ the Hamiltonian $H$
can be written as

\begin{equation}
H=ghg^{-1},\qquad hP_{N}=E_{N}P_{N}  \label{6.4}
\end{equation}
with 
\begin{equation}
h=a_{ik}^{\alpha }T_{ik}^{\alpha }+a_{ik,lm}^{\alpha \beta }T_{ik}^{\alpha
}T_{lm}^{\beta }  \label{6.5}
\end{equation}
where $a_{ik}^{\alpha }$ and $a_{ik,lm}^{\alpha \beta }\,$are constants
(subject to some further conditions \cite{GLKO}) and 
\begin{equation}
T_{ik}^{\alpha }=s_{i}^{\alpha }\partial _{s_{k}},\qquad \alpha =0,1,\qquad
i,k=1,...,n.  \label{6.6}
\end{equation}
\end{enumerate}

In other words, the gauge rotated Hamiltonian $h\,$is an element of the
enveloping algebra of an affine Lie algebra $aff\left( n,{\Bbb {R}}\right) $
(or one of its subalgebras). It is clear that (\ref{6.5}) guarantees that
the Hamiltonian $H$ will preserve, or decrease the order of the polynomials $%
P_{N}$. This is a concrete realization of the flag condition (\ref{6.2}).

We mention that all known quadratically superintegrable systems are exactly
solvable, in particular those of Section 4 \cite{TTW}. For the generalized
harmonic oscillators the gauge factor $g$ is equal to the ground state wave
function and is energy independent. The generalized Coulomb systems have
been reduced to the harmonic oscillators ones (see (\ref{5.16}) and (\ref
{5.18})). However, due to the interchange of the energy and the coupling
constant, the gauge factor $g$ will be energy dependent.

The aim of this Section is to show how exact solvability manifests itself in
discrete space--time. First of all, let us consider an arbitrary
one--dimensional linear spectral problem 
\begin{equation}
L\left( \partial _{x},x\right) \psi \left( x\right) =\lambda \psi \left(
x\right) .  \label{6.7}
\end{equation}
Let $x=0$ be a regular point of this equation. Then any solution can be
expanded into a Taylor series 
\begin{equation}
\psi \left( x\right) =\sum_{k=0}^{\infty }a_{k}x^{k}.  \label{6.8}
\end{equation}
Using the umbral correspondence we write the umbral equation 
\begin{equation}
L\left( \Delta ,x\beta \right) \psi \left( x\beta \right) =\lambda \psi
\left( x\beta \right)   \label{6.9}
\end{equation}
with the same eigenvalue $\lambda \,$as in the ODE (\ref{6.7}). Viewed as a
difference equation, eq. (\ref{6.9}) will have a formal power series
solution 
\begin{equation}
\psi \left( x\beta \right) 1=\sum_{k=0}^{\infty }a_{k}^{k}\left( x\beta
\right) ^{k}\cdot 1.  \label{6.10}
\end{equation}
In particular, if (\ref{6.8}) is a polynomial solution, then eq. (\ref{6.10}%
) will also be a finite sum of terms involving the basic polynomials of the
operator $\Delta $. Thus, (\ref{6.10}) will also be a polynomial and all
convergence problems disappear.

Now let us turn to the specific case of the generalized harmonic oscillator
system with Hamiltonian $H_{I}$ (see eq. \ref{5.12}) and its discretization (%
\ref{5.27}). The gauge factor $g$ of eq. (\ref{6.3}) and (\ref{6.4}) is 
\begin{equation}
g=x^{p_{1}}y^{p_{2}}\exp \left[ -\frac{\omega \left( x^{2}+y^{2}\right) }{2}%
\right] ,\qquad a=p_{1}\left( p_{1}-1\right) ,\qquad b=p_{2}\left(
p_{2}-1\right) .  \label{6.11}
\end{equation}
We put $\omega x^{2}=s_{1},$ $\omega y^{2}=s_{2}\,$ and in these variables
the $aff\left( 2,{\Bbb {R}}\right) \,$operators of eq. (\ref{6.6}) reduce to

\[
J_{1}=\partial _{s_{1}},\quad J_{2}=\partial _{s_{2}},\quad
J_{3}=s_{1}\partial _{s_{1}},\quad J_{4}=s_{2}\partial _{s_{2}},
\]
\begin{equation}
J_{5}=s_{2}\partial _{s_{1}},\quad J_{6}=s_{1}\partial _{s_{2}}.
\label{6.11b}
\end{equation}
The gauge rotated hamiltonian $h$ and gauge rotated integrals of motion $%
\widehat{x_{1}}=g\widehat{X_{1}}g^{-1}$, $\widehat{x_{2}}=g\widehat{X_{2}}%
g^{-1}\,$can now be written as \cite{TTW} 
\[
h=-2J_{3}J_{1}-2J_{4}J_{2}+2J_{3}+2J_{4}-\left( 2p_{1}+1\right) J_{1}-\left(
2p_{2}+1\right) J_{2}
\]
\[
\widehat{x_{1}}=2J_{3}J_{1}-2J_{4}J_{2}-2J_{3}+2J_{4}+\left( 2p_{1}+1\right)
J_{1}-\left( 2p_{2}+1\right) J_{2}
\]
\[
\widehat{x_{2}}=4J_{3}J_{5}+4J_{4}J_{6}-8J_{3}J_{4}+2\left( 2p_{1}+1\right)
J_{5}-2\left( 2p_{2}+1\right) J_{3}+
\]
\begin{equation}
-2\left( 2p_{1}+1\right) J_{4}+2\left( 2p_{2}+1\right) J_{6}.  \label{6.12}
\end{equation}
By construction\smallskip , all three of these operators will conserve the
flag of polynomials 
\begin{equation}
P_{n}\left( s_{1},s_{2}\right) =\langle \left( s_{1}\right) ^{N_{1}}\left(
s_{2}\right) ^{N_{2}}\mid 0\leq N_{1}+N_{2}\leq n\rangle   \label{6.13}
\end{equation}
and this is the reason why the superintegrable system with Hamiltonian $%
H_{I}\,$is exactly solvable. The actual solutions of eq. (\ref{6.4})\ are
Laguerre polynomyals:

\begin{equation}
HP_{nm}=E_{nm}P_{nm},\qquad E_{mn}=n+m,  \label{6.14}
\end{equation}

\[
\qquad P_{nm}\left( x,y\right) =L_{n}^{\left( -1/2+p_{1}\right) }\left(
\omega x^{2}\right) L_{m}^{\left( -1/2+p_{2}\right) }\left( \omega
y^{2}\right) . 
\]

The umbral discretization will preserve the above properties and will give
umbral Laguerre polynomial expressed in terms of $x\beta _{x}$ and $y\beta
_{y}$. The algebra $aff\left( 2,{\Bbb {R}}\right) \,$is represented by
difference operators 
\[
\widetilde{J_{1}}=\Delta _{s_{1}},\quad \widetilde{J_{2}}=\Delta
_{s_{2}},\quad \widetilde{J_{3}}=\left( s_{1}\beta _{1}\right) \Delta
_{s_{1}},\quad \widetilde{J_{4}}=\left( s_{2}\beta _{2}\right) \Delta
_{s_{2}}, 
\]
\begin{equation}
\widetilde{J_{5}}=\left( s_{2}\beta _{2}\right) \Delta _{s_{1}},\qquad 
\widetilde{J_{6}}=\left( s_{1}\beta _{1}\right) \Delta _{s_{2}}.
\label{6.15}
\end{equation}
The formulas (\ref{6.12}) remain the same (with $J_{i}\rightarrow \widetilde{%
J_{i}}$) and all commutation relations are preserved, as are polynomial
solutions. For similar results formulated in terms of operators acting in
Fock spaces and the notion of isospectral discretization see Turbiner et al.
(\cite{T1}--\cite{T3}), and \cite{DMS} for further discussions.

\section{Conclusions}

Much, if not all of nonrelativistic quantum mechanics can be viewed as the
''theory of the enveloping algebra of the Heisenberg algebra''.

Indeed, let us define the Heisenberg algebra $H_{n}$ by the relations 
\begin{equation}
\left[ X_{j},Y_{k}\right] =\delta _{jk}C\qquad j,k=1,...,n  \label{7.1}
\end{equation}
and then put 
\begin{equation}
X_{j}=x_{j},\qquad Y_{k}=-i\hbar \partial _{x_{k}},\qquad C=i\hbar .
\label{7.2}
\end{equation}

We can say that all quantum mechanical operators lie in the enveloping
algebra of $H_{n}$, or in an extension of the enveloping algebra obtained by
adding all formal power series in $x_{1},...,x_{n}$, $p_{1},...,p_{n}$.

If we replace the coordinates $x_{j}$ and the momenta $p_{j}\,$by some other
quantities satisfying the relations (\ref{7.1}) then all polynomials and all
power series in these objects will commute in the same way as the
corresponding quantum mechanical quantities.

Indeed, the umbral correspondence $x_{i}\rightarrow x_{i}\beta _{i}$, $%
\partial _{x_{i}}\rightarrow \Delta _{x_{i}}\,$preserves the commutation
relations (\ref{7.1}) between quantum mechanical operators. Thus, the umbral
correspondence allows us to consider quantum mechanics on a lattice and to
preserve all properties of quantum mechanics in continuous space and time
that are expressed in terms of the commutation properties of physical
quantities (quantum mechanical operators). In particular, infinitesimal
point symmetries are preserved, as shown in Sections 3 and 4, respectively.
Exact solvability is reduced to an algebraic property and then discretized
in Section 5 (and also in the papers \cite{ST}, \cite{T25}--\cite{T3}).

Some nonalgebraic properties are lost in the discretization. For instance,
the ''umbral vector fields'' (\ref{2.32}), obtained by the umbral
correspondence, do not generate global transformations (like rotations, or
dilations). The solutions of umbral equations (obtained by the umbral
correspondence) are often formal, i.e. they may diverge.

The physical content of this article is based on the results contained in
Section 2, where we presented and proved several theorems that, to our
knowledge, extend the previously known umbral formalism to linear difference
equations. In particular we associate with a linear differential equation an
abstract operator equation, written in terms of delta operators, using the
umbral correspondence (\ref{2.26}). Any representation of $\Delta $ and $%
\beta $ in terms of shift--invariant operators provides a difference
equation, whose analytic solutions can be obtained from the solutions $%
\widehat{f}$ of the operator equation (\ref{2.33}) via the projection $%
\widehat{f}\cdot 1=f\left( x\beta \right) \cdot 1$. These are the umbral
solutions admitted by a given difference equation. The other possible
solutions do not have a continuous limit and are not provided by the umbral
approach.

Many of the results presented here can be considered also in the case of $q$%
--difference operators \cite{LNO3}. The delta operator $U$ in this case is
defined in terms of a $q$--shift operator satisfying 
\begin{equation}
T_{q}f\left( x\right) =f\left( qx\right) .  \label{7.3}
\end{equation}
In the simplest case, the $q$--difference operator reads 
\begin{equation}
\Delta _{q}=\frac{1}{\left( q-1\right) x}\left( T_{q}-1\right) ,\qquad
\lim_{q\rightarrow 1}\Delta _{q}f=f_{x}  \label{7.4}
\end{equation}
and its conjugate operator, obtained imposing the Heisenberg commutation
relation 
\begin{equation}
\left[ \Delta _{q},x\beta _{q}\right] =1  \label{7.5}
\end{equation}
is written in terms of a $q$--shift and differential operators as 
\begin{equation}
\beta _{q}=\left( q-1\right) \left( qT-1\right) ^{-1}x\partial _{x}.
\label{7.6}
\end{equation}

In this case we can still consider the umbral correspondence, but the $%
\Delta _{q}\,$operator is not a shift invariant operator, as $\left[ \Delta
_{q},T_{q}\right] \neq 0$.

Among open questions, presently under consideration, we mention the
following.

\begin{itemize}
\item  The umbral correspondence has lead us to various linear umbral
equations and difference equations. It would be of considerable interest to
study their solutions directly and in the case of polynomial solutions,
establish their relation to orthogonal polynomials of discrete variables,
known in the literature \cite{Riord}, \cite{NSU}.

\item  Although many of the results presented in this paper can also be
carried out in the $q$--difference case \cite{LNO3}, \cite{KC}, \cite{LNO2},
a reformulation of the umbral theory in this case is necessary.

\item  The simultaneous diagonalization of commuting sets of second order
differential operatores is intimately related to the separation of variables
in the Schr\"{o}dinger equation. It would be important to investigate common
solutions of commuting sets of difference operators from this point of view.

\item  A related problem is that of establishing a connection between umbral
formalisms introduced in different coordinate systems, e.g. the cartesian
quantities $\Delta _{x}$, $\Delta _{y}$, $\beta _{x}$, $\beta _{y}$ and the
corresponding polar ones, or parabolic ones $\Delta _{\xi }$, $\Delta _{\eta
}$, $\beta _{\xi }$, $\beta _{\eta }\,$introduced in Sections 4 and 5.
\end{itemize}

{\bf Acknowledgements\newline
}We thank J. Negro, M. del Olmo, M. A. Rodriguez, J. Szmigielski and A.
Turbiner for helpful discussions. P. W. research was partly supported by
research grants from NSERC of Canada and FQRNT du Quebec. P. T. and P. W.
thank the Universit\`{a} Roma Tre for its hospitality and support, similarly
D. L. thanks the CRM. P. T. benefitted from a CRM--ISM\ fellowship. A NATO
collaborative grant n. PST.CLG.978431 is gratefully acknowledged.

\end{document}